\documentclass[pra,twocolumn,superscriptaddress,floatfix,nofootinbib,amsmath,amssymb]{revtex4-1}

\usepackage[english]{babel}
\usepackage{float}
\usepackage{graphicx}
\usepackage{bm}
\usepackage{braket}
\usepackage{xcolor}
\usepackage{amsfonts}
\usepackage{comment}
\usepackage{hyperref}
\usepackage{cleveref}
\usepackage{url}
\usepackage{bbold}
\usepackage{realboxes}

\usepackage{enumitem}
\usepackage{csquotes}
\usepackage[boxruled]{algorithm2e}

\usepackage{amsmath}
\DeclareMathOperator\arctanh{arctanh}

\newcommand{\Nres}{N_\text{res}}
\newcommand{\Mres}{M_\text{res}}

\definecolor{bkgd}{RGB}{240,242,246}
\definecolor{ceruleanblue}{rgb}{0.16, 0.32, 0.75}
\definecolor{orange-red}{rgb}{1.0, 0.27, 0.0}
\definecolor{anotherblue}{RGB}{37,92,243}
\definecolor{blackblue}{RGB}{46,60,85}
\definecolor{goldyellow}{RGB}{199,146,12}
\begin{document}

\title{Nonequilibrium Monte Carlo for unfreezing variables \\ in hard combinatorial optimization }

\affiliation{Google Quantum AI, Venice, CA 90291
}

\affiliation{Google, Mountain View, CA
}

\affiliation{Dipartimento di Fisica, Sapienza Università di Roma, P.le Aldo Moro 5, 00185 Rome, Italy}

\affiliation{Google Quantum AI, Zurich, Switzerland
}

\author{Masoud Mohseni}
\email{mohseni@google.com}
\affiliation{Google Quantum AI, Venice, CA 90291
}

\author{Daniel Eppens}
\affiliation{Google Quantum AI, Venice, CA 90291
}

\author{Johan Strumpfer}
\affiliation{Google, Mountain View, CA
}

\author{Raffaele Marino}
\affiliation{Dipartimento di Fisica, Sapienza Università di Roma, P.le Aldo Moro 5, 00185 Rome, Italy}

\author{Vasil Denchev}
\affiliation{Google Quantum AI, Venice, CA 90291
}

\author{Alan K. Ho}
\affiliation{Google Quantum AI, Venice, CA 90291
}

\author{Sergei V. Isakov}
\affiliation{Google Quantum AI, Zurich, Switzerland
}

\author{Sergio Boixo}
\affiliation{Google Quantum AI, Venice, CA 90291
}

\author{Federico Ricci-Tersenghi}
\email{federico.ricci@roma1.infn.it}
\affiliation{Dipartimento di Fisica, Sapienza Università di Roma, P.le Aldo Moro 5, 00185 Rome, Italy}
\affiliation{CNR, Nanotec, and INFN, Sezione di Roma I, P.le Aldo Moro 5, 00185 Rome, Italy}

\author{Hartmut Neven}
\affiliation{Google Quantum AI, Venice, CA 90291
}

\date{\today}


\begin{abstract}

Optimizing highly complex cost/energy functions over discrete variables is at the heart of many open problems across different scientific disciplines and industries. A major obstacle is the emergence of many-body effects among certain subsets of variables in hard instances leading to critical slowing down or collective freezing for known stochastic local search strategies. An exponential computational effort is generally required to unfreeze such variables and explore other unseen regions of the configuration space. Here, we introduce a quantum-inspired family of nonlocal Nonequilibrium Monte Carlo (NMC) algorithms by developing an adaptive gradient-free strategy that can efficiently learn key instance-wise geometrical features of the cost function. That information is employed on-the-fly to construct spatially inhomogeneous thermal fluctuations for collectively unfreezing variables at various length scales, circumventing costly exploration versus exploitation trade-offs. We apply our algorithm to two of the most challenging combinatorial optimization problems: random k-satisfiability (k-SAT) near the computational phase transitions and Quadratic Assignment Problems (QAP). We observe significant speedup and robustness over both specialized deterministic solvers and generic stochastic solvers. In particular, for 90\% of random 4-SAT instances we find solutions that are inaccessible for the best specialized deterministic algorithm known as Survey Propagation (SP) with an order of magnitude improvement in the quality of solutions for the hardest 10\% instances. We also demonstrate two orders of magnitude improvement in time-to-solution over the state-of-the-art generic stochastic solver known as Adaptive Parallel Tempering (APT).

\end{abstract}
\maketitle

Over the past few decades there has been a growing interest in establishing connections between the concepts and tools of statistical physics and computer science. Notably, spin-glasses provide a universal language for representing computational or learning tasks over discrete variables \cite{MezardBook,nishimori_statistical_2001}. The hardness of approximating combinatorial optimization problems or probabilistic inference in graphical models can be mapped to difficulties in evaluating marginal probabilities,  estimating the partition functions, or sampling over the Boltzmann distributions for low energy states of spin-glass systems \cite{MezardBook,moore_nature_2011}. These are computational bottlenecks that appear in a wide range of applications including training and inference in energy-based models \cite{LeCun06tutorial}, structured input/output machine learning \cite{goodfellow_deep_2016}, Bayesian learning \cite{bickel_bayesian_1996}, and causal inference  \cite{peters_elements_2017}. Moreover, the nonequilibrium dynamics of spin-glass systems and their metastable states represent steady-state attractors in dynamical systems \cite{hopfield_neural_1982}, associative memory \cite{hopfield_neural_1982, nishimori_statistical_2001}, and storage capacity of classical and quantum neural networks \cite{Gardner_1988,Lewenstein_2021}. Many of these outstanding open problems can be reformulated as disentangling or learning correlations in many-body interacting systems. Consequently, there is a significant opportunity for developing physics-based solvers and models to compute or learn such correlations. 

Historically, several important deterministic algorithms for approximating the partition function, or evaluating marginal probability distributions, have been physics-inspired or have direct physical correspondence; in particular Replica Symmetry Breaking (RSB) and the cavity method \cite{mezard1987spin,MezardBook}, belief propagation algorithms \cite{BP03}, and tensor-network contractions \cite{ran_tensor_2020,rams_tensors_2021}. Advanced concepts and tools, such as 1RSB cavity methods have lead to certain generalization of belief propagation techniques known as Survey Propagation (SP) which performs accurately over problems with locally tree-like graphs \cite{mezard_analytic_2002, maneva_new_2007, marino_backtracking_2016}. A general probabilistic physics-inspired approach for sampling that can be applied to problems with discrete or continuous variables is Markov Chain Monte Carlo (MCMC) by leveraging local thermal fluctuations enforced by Metropolis-Hastings updates~\cite{Metropolis49,Hastings70}. This class includes Simulated Annealing~\cite{Kirkpatrick671}, Replica-exchange Monte Carlo or Parallel Tempering (PT) ~\cite{PTreview}, Langevin Monte Carlo \cite{parisi_correlation_1981}, and Hamiltonian Monte Carlo \cite{hoffman_no-u-turn_2011}. 

Despite this progress, one of the main challenges is the exponentially slow mixing of local equilibrium dynamics of MCMC sampling for problems with multimodal distributions. To tackle this deficiency, one typically employs advanced techniques which combine various cluster update strategies over a baseline MCMC algorithm. This
includes Swendsen-Wang-Wolf cluster updates~\cite{Swendsen_Wang87,Wolf89}, Houdayer or Iso-energetic cluster   moves~\cite{Houdayer2001,Katzgraber_ICM_2015}, or Hamze-Freitas-Selbey algorithm ~\cite{HF04,Selby14,Hen_2017}. %
However, these approaches either break down for frustrated systems ~\cite{Wolf89}, or percolate for systems with dimensions $D>2$ \cite{Houdayer2001}. Other cluster update techniques invoke randomly selected tree-like subgraphs for efficient sampling with dynamic programming ~\cite{HF04,Selby14,Hen_2017}. However, such clusters are not necessarily related to the actual structures, or backbones \cite{maneva_new_2007}, of the underlying problems. Another class of nonlocal physics-based approaches relies on quantum fluctuations to induce cluster updates such as quantum annealing or adiabatic quantum computation~\cite{Lidar18}, dissipative quantum tunneling~\cite{Boixo16},
coherent many-body delocalization ~\cite{Kechedzhi18}, or shallow depth quantum circuits \cite{mcclean_low-depth_2021}. However, the potential computational
power of quantum computers over classical computation
is yet not well understood~\cite{troyer_speedup_science_2015,Mohseni17} as they could suffer from  decoherence effects, finite control precision, sparse and low-dimensional underlying graphs, significant embedding overheads, Griffiths singularities, and typically exponentially vanishing quantum Hamiltonian gaps. Nevertheless, some of these limitations could be partially mitigated by invoking alternative or complementary physical mechanisms \cite{mcclean_low-depth_2021}; e.g., by inhomogeneous nonequilibrium quantum annealing schedules~\cite{mohseni_engineering_2018} or hybrid quantum-assisted PT~\cite{Mohseni_patent}.

In this work, we demonstrate that nonlocal quasi-equilibrium cluster updates can be constructed fully classically by iteratively computing the local marginals and higher-order correlations of discrete variables during the actual runtime of Monte Carlo sampling. 
We introduce a new family of algorithms with subroutines that have tunable local temperatures for key subset of variables, which we denote as \enquote{\textit{surrogate backbones}}, that are learned in a instance-wise fashion. The surrogate backbones consist of the variables that all hold same values over all the high-quality solutions in a given basin of attraction. This allows us to optimize separately for exploration and exploitation subroutines and create non-trivial interplay between these two mechanisms. This is in contrast to generic MCMC-based heuristic solvers, such as SA and PT, that typically invoke global temperatures for replicas and which have to be simultaneously optimized for both exploration (overcoming large energy barriers) and exploitation (local searching within each basin of attractions) leading to unavoidable computational trade-offs.

Specifically, we first control the locality of our search by building an adjustable localized  surrogate Hamiltonian for each replica to pin them to a particular basin of attraction in a given energy-scale. This allows us to reliably use loopy belief propagation, even for problems with arbitrary graph dimension, to estimate local fields and correlation functions. Using that information, we grow clusters of highly rigid variables in each basin of attraction. Each of these clusters act as an ansatz for the backbone of the surrogate localized Hamiltonian. These backbones reveal the essential geometrical features of the loss function. Subsequently, we construct inhomogeneous temperature profiles across each replica for efficient exploration. This is achieved by significantly boosting the temperature of each backbone ansatz. Finally, we devise frequent unlearning phases by employing standard (local and homogeneous) replica-exchange Monte Carlo. This phase is inspired by unlearning or negative phase in Boltzmann machines \cite{goodfellow_deep_2016}. The homogeneity is mainly inserted as an important mechanism to mitigate inductive bias. Here, inductive bias is physically manifested as accumulation of domain walls or topological defects at the boundaries of our backbone ansatz and the rest of variables. We iteratively alternate between these three subroutines in a hierarchical fashion across many replicas that are adaptively placed near the spin-glass phase transition. In other words, our algorithm respects the natural inhomogeneity of the problem and tackles the exponentially slowing down of MCMC sampling with inhomogeneous control of the energy/time scale separation for the rigid or frozen variables. 

Our approach does not make any assumptions about the nature of two or higher-body interactions among variables, distribution of couplings, graph connectivity, or dimension of the problem and thus can be applied as a generic solver to a wide variety of problem classes. We observe orders of magnitude performance improvements for a number of NP-hard problems including random 4-SAT problems consisting of 5000 variables with clause to variable ratio of 9.884 which is past the estimated rigidity threshold and very near the computational phase transition. By introducing a generalization of the whitening procedure \cite{parisi_local_2005, parisi_survey-propagation_2008, maneva_new_2007}, we find several independent high quality solutions for the hardest $4$-SAT instances that contain large frozen backbones of size $O(N)$. This task is generally believed to be exponentially hard to achieve with local solvers for sufficiently low-energy states of hard instances that are deep in the frozen regime \cite{moore_nature_2011,marino_backtracking_2016}. Some of these frozen solutions could not be found with even $O(1000)$ repetitions of standard APT algorithm. We use the complexity of cluster of solutions for 4-SAT formulas to provide a measure of instance-wise hardness. We observe large fluctuations for Backtracking Survey Propagation (BSP) and small fluctuations for NMC over such hard instances, making the latter a much more reliable solver.

%%%%%%%%%%%%%%%%%%%%%%%%%%%%%%%%%%%%%%%%%%%%%%%%%%%%%%%%%%%%%%%%%%%%%%%%%
\begin{figure*}[ht]
    \centering
    \includegraphics[width=\textwidth]{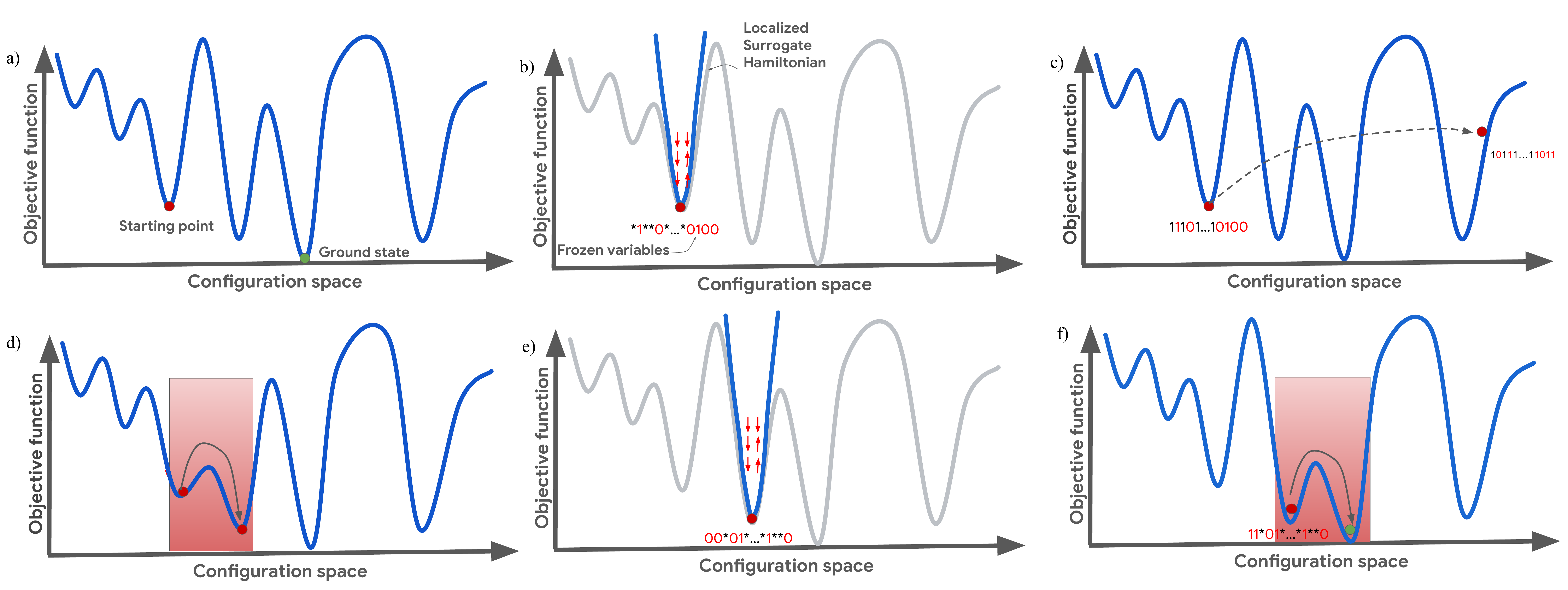}
    \caption{Schematic diagram of Nonequilibrium Monte Carlo: (a) Generate seed solution using an off-the-shelf solver; (b) Build localized surrogate problems with local penalty terms to Hamiltonian, and then perform efficient inference to estimate first and higher order marginals, and finally grow surrogate backbone; Using the information provided in (b) to perform either (c) and/or (d); (c) Implement a nonlocal full spin-flip of backbones; (d) Construct inhomogeneous Monte Carlo by significantly boosting the temperatures for backbone variables, essentially flattening the relevant energy barrier; (e) and (f) Repeat the procedure until finding the ground state or another high quality solution in the target approximation ratio.}
    \label{fig:NMC_animation}
\end{figure*}
%%%%%%%%%%%%%%%%%%%%%%%%%%%%%%%%%%%%%%%%%%%%%%%%5%%%%%%%%%%%%%%%%%%%%%%%%%

\section{Nonequilibrium Nonlocal Monte Carlo}

\label{sec:MCMClimitations}

The common picture for a complex energy landscape is that of a function defined in a very high-dimensional space with a large number of local minima and large barriers between them. The computational complexity in sampling such a corrugated landscape comes from the conflicting needs of visiting low-energy minima, while simultaneously being able to overcome high barriers.
It is worth mentioning that often these complex landscapes in high dimensional spaces also present entropic barriers that affect both classical and quantum algorithms \cite{bellitti_entropic_2021}. However at low enough temperatures the energetic barriers are the main obstacle.

The most widely used algorithms for performing the sampling of a complex energy landscape are based on MCMC. However, standard MCMC where the temperature of the bath is kept constant to $T$ is deemed to fail: large values of $T$ are required for jumping over large barriers, but small values of $T$ are required to visit low-energy configurations, thus trapping the evolution of the system in some local minima.

A straightforward approach to this problem is to allow the temperature to change during the simulation. If one is interested in finding just one low energy configuration in a \emph{optimization problem} then the use of Simulated Annealing \cite{Kirkpatrick671} where the temperature is gradually decreased during the simulation may be of great help.
However, if one is interested in the \emph{sampling problem}, many different low energy minima must be visited by the algorithm and thus the temperature needs to be raised and lowered back again many times. This is the idea behind replica-exchange MC or parallel tempering  which is currently the best general purpose algorithm for sampling complex energy landscapes \cite{PTreview, katzgraber_introduction_2011}.

Nevertheless all the above algorithms have a strong limitation: they use the same global temperature for updating each microscopic variable of the system under study. In other words, the temperature is constant over the entire system. This is required if one wants to sample from the Gibbs-Boltzmann distribution at a given temperature.
However, if the main aim of the simulation is to bring the temperature sufficiently close to zero to eventually sample from the many low energy states, then it is not clear why one should keep the temperature uniformly constant over the entire system, as the system response to temperature is not uniform. There might be alternative inhomogeneous schemes for updating the temperatures. Generally, it is not obvious why changing the temperature from region to region of the system could significantly help navigating between low energy minima. We provide an intuitive argument for these inhomogeneous temperature profiles before developing our algorithm and showing convincing numerical evidence.

Very strong heterogeneities are common in disordered and frustrated systems \cite{glotzer1998dynamical,banos2010static}. For typical configurations obtained by sampling at a given low temperature, there are regions where the interactions are mostly satisfied and thus variables are very rigid (almost frozen); while, there are other regions where interactions are much less satisfied, and consequently the variables are less constrained and can vary more easily \cite{lage2014message}. This strong heterogeneity in the rigidity of different parts of the system under study produces very different time scales in its evolution \cite{ricci2000glassy}.

For simplicity let us assume the system can be decomposed in two parts or regions: a more rigid and a more floppy region. In order to optimize the system, one needs to bring the temperature low enough, but at such low temperature the more rigid part is completely frozen and does not evolve at all. On the contrary, when the temperature is raised high enough to update the more rigid part, any correlation in the less rigid part is completely washed out and the optimization process on that part will need to be restarted from scratch. This is the problem when using any uniform or homogeneous temperature changing protocol on a very heterogeneous system: to update the most rigid parts of the system, the algorithm must increase the temperature globally and so forgets any good correlation that has developed in the less rigid part of the system.

Starting from this observation, our idea is to use different temperatures in different parts of the system. In this way one can update the most rigid parts of a system without destroying the correlations that have been developed in the least rigid part.
These nonuniform updates would violate detailed balance, so it can not be used as a dominating mechanism for fair sampling at a non-zero temperature over the microscopic degrees of freedom. However, as we show below, when invoked occasionally in conjunction with standard MCMC, they lead to a nonequilibrium steady state with an effective balance condition that samples from the low-energy states. Moreover, in the optimization problems and in sampling at $T=0$, the aim is to find one (or many) lowest energy configurations. In this case the heuristic algorithm based on the idea of using different temperatures in different parts of the system is fine as long as all temperatures are eventually made sufficiently small.

In terms of the corrugated energy landscape, our aim is to move between low-energy minima without bringing the entire system to a high energy above the barriers; something that in principle could be achieved by quantum tunneling. Here, however, our idea is to implement classical cluster moves where only very rigid variables are given a larger thermal (i.e.\ stochastic) energy. The rationale beyond this choice is the following: in a low-energy minimum where variables have different levels of rigidity, i.e.\ very different correlations among variables, the curvature of the landscape strongly depends on the direction, i.e.\ on the subset of variables that are flipping at each step of the algorithm. For instance, flipping a very correlated set of variables could significantly increase the energy, thus it corresponds to climbing up an energy barrier.
By coupling only this subset of variables to a high temperature bath we are effectively lowering the barrier, which facilitate transitions between different low-energy minima. In principle, we could flip all of the correlated variables at once, which would be more effective when there is inherent $\mathbb{Z}_2$ symmetry, 
and subsequently boost their local temperature.  Overall, in contrast with an algorithm where the temperature is raised everywhere, here the minima that we are trying to connect are still well defined thanks to the fact that the majority of variables are still coupled to a bath with a very low temperature. 

\begin{figure*}[ht]
    \centering
    \includegraphics[width=\textwidth]{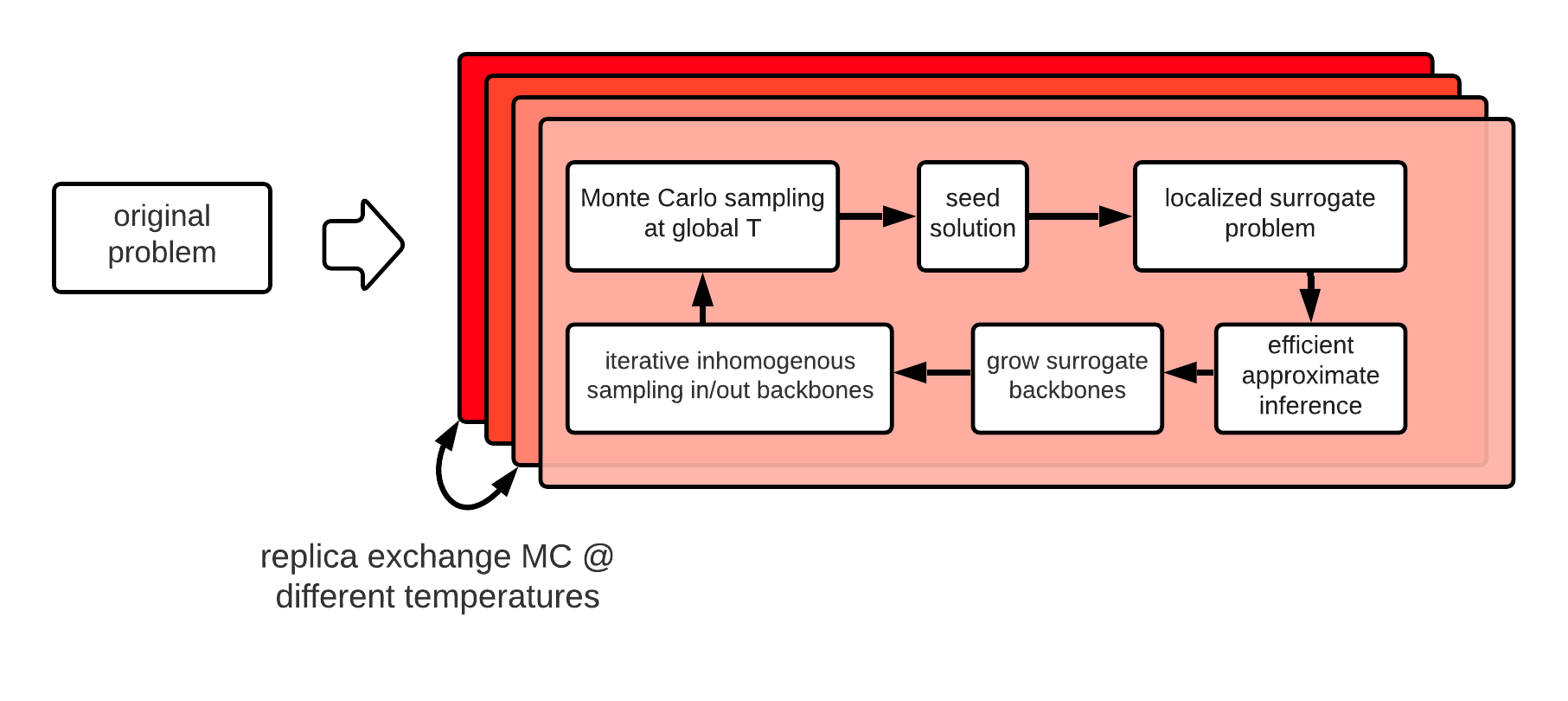}
    \caption{Outline of NMC algorithm for a given replica at temperature T embedded in a parallel tempering or replica-exchange MC algorithm. An approximate solution (S*) from a standard MCMC is used to build localized surrogate Hamiltonians for a given replica. The surrogate problems can be efficiently sampled with Loopy Belief Propagation (LBP) to calculate local marginals and higher-order correlation functions. Using LBP, the ansatz backbone of surrogate Hamiltonians are evaluated which could act as shortcuts for the original problem. Nonequilibrium  MCMC is invoked iteratively with a nonuniform temperature profile in each cycle: first temperatures are significantly boosted inside the backbones to find their low energy states efficiently while non-backbone variables are fixed to S*. This is followed by efficient sampling of non-backbone variables that are conditioned on a given low-energy state of the backbone. This process is repeated many times in each cycle and can be applied at different replicas with different base temperatures within a replica-exchange MC algorithm.}
    \label{fig:schematic_nonlocalMCMC}
\end{figure*}

We overcome the failures of local and homogeneous Monte Carlo sampling by iterating between subroutines that are customized towards exploitation and exploration of the energy landscape.  To construct such paradigm of computation, however, there are several major outstanding challenges and open questions: how can we actually compute or learn such elusive subsets of  rigid variables in a given basin of attraction for strongly disordered and frustrated systems? How can we use such information to grow meaningful clusters? And ultimately how can we create the desired nonlocal moves? We address all these questions in the subsequent sections. 
Motivated by the phenomenological description of local homogeneous stochastic search strategies presented in this section, we first provide a high-level and intuitive illustration of the algorithm in Fig. \ref{fig:NMC_animation}. A particular realization of NMC that is build on top of an APT framework is presented in Algorithm \ref{algorithm} and Fig. \ref{fig:schematic_nonlocalMCMC}. For a short description of our APT algorithm see App. \ref{APT}. In the next section, we provide a detailed construction of our algorithm.

\bigskip

\begin{minipage}{1\linewidth}
\begin{algorithm}[H]
\SetAlgoLined
\SetInd{1em}{1em}

\While{good solutions not found}{

\textbf{Replica exchange MC}: Adaptive homogeneous replica-exchange MC on the entire problem.

\For{replicas at low temperatures}{

\textbf{Generate seeds}: Find a low energy state as a seed solution, $s^{*}$.

\For{each seed}{

\textbf{Build localized problems}: Construct a localized surrogate Hamiltonian around the neighborhood of a seed solution.

\textbf{Infer correlations}: Use efficient approximate inference techniques, such as LBP, to estimate marginals over the localized surrogate problem.

 \textbf{Grow backbones}: Threshold the correlations to construct surrogate backbones over rigid variables.

}

\While{arriving at a steady state}{

\textbf{Nonlocal exploration:} Inhomogeneous Monte Carlo on backbone subproblem by conditioning over non-backbone variables.

\textbf{Local exploitation:} 
Inhomogeneous Monte Carlo on non-backbone subproblem by conditioning over backbone variables.

\textbf{Unlearning phase:} Perform homogeneous MCMC sampling on full problem to repair topological defects at the backbone boundaries.

}

}

} 
 \caption{ \textsc{Nonequilibrium Monte Carlo} (NMC)}
 
 \label{algorithm}
\end{algorithm}
\end{minipage}

%%%%%%%%%%%%%%%%%%%%%%%%%%%%%%%%%%%%%%%%%%%%%%%%%%%%%%%%%%%%%%%%%%%%%%%%%
%%%%%%%%%%%%%%%%%%%%%%%%%%%%%%%%%%%%%%%%%%%%%%%%5%%%%%%%%%%%%%%%%%%%%%%%%%

\bigskip\bigskip\bigskip

\section{Constructing localized surrogate Hamiltonians}

\label{sec:H_surrogate}

We are interested in finding many different low energy configurations for an energy function $H(\textbf{s}): \{-1,1\}^N \to \mathbb{R}$. Let's consider a generalized spin-glass system including interactions up to the $k$-th order: 
\begin{equation} \label{eq:Hamiltonian}
\begin{split}
H(\textbf{s}) = -\sum_i h_i s_{i} - \sum_{(ij)\in E} J_{ij} s_i s_j- \dots \\ -\sum_{(ij \dots k)\in E} J_{ij\dots k} s_i s_j \dots s_k \;,
\end{split}
\end{equation}
where $E$ is the edge-set of the interactions in the hypergraph. For simplicity, in this section we first build our algorithm for systems with pairwise interactions. We will provide the generalization to high order interactions in Sec. \ref{sec:LBPhigher}.

Let's start with running any state-of-the-art Monte Carlo sampling techniques, such as replica-exchange MC or parallel tempering (PT) \cite{katzgraber_introduction_2011},  to arrive at a fairly high-quality spin configuration $\textbf{s}^{\star}=\{\textbf{s}^{\star}_i\}_{i,\ldots,N} \in \{-1,1\}^N$ at a given replica. If the running time of PT is sufficiently long, then the state $\textbf{s}^{\star}$ will likely be a low-energy configuration close to a minimum of the energy landscape and departing from such configuration $\textbf{s}^{\star}$ would be difficult with PT and practically impossible for Monte Carlo replicas at fixed low temperatures. Even if $H(\textbf{s}^{\star})$ is low enough to satisfy our goals, it is often useful to find a {\it diversity} of configurations of the same low energy. Recently, a notion of diversity measure for spin glasses has been introduced which can be enhanced for low-dimensional systems using inhomogeneous quantum annealing schedules guided by approximate tensor-network contraction preprocessing \cite{mohseni_diversity_2021}. However, there is no known technique for how to enhance the diversity of solutions over general problems living over arbitrary hypergraphs.  To this end, we would like to propose nonlocal \textit{cluster} moves involving a large number of spin variables. It is known that cluster algorithms for strongly disordered systems do not work because strong correlations (used to define clusters) percolate at a wide range of temperatures \cite{zhu_efficient_2015}. To avoid percolation, we resort to heuristic approaches that might not satisfy the 
detailed balance condition, but could be very effective if the proposed change has $\Delta H(\textbf{s},\textbf{s}^{\prime}) = O(1)$ with Hamming distance $D_H(\textbf{s},\textbf{s}^{\prime})=\sum_i (1-\delta_{s_{i},s_{i^\prime}})=O(N)$. 

The main idea of this work is to compute the local properties of the energy landscape and use that information to build nonlocal moves in the configuration space. It should be noted that we are in a discrete space and we cannot simply compute derivatives to get an idea of local geometry, thus we need to estimate the corresponding measures that are  first and second order marginals, namely magnetizations and correlations. Unfortunately, standard MCMC sampling schemes \cite{katzgraber_introduction_2011} are not reliable to provide good estimates of such local quantities as their dynamics are designed to recover ergodicity, and thus often wander among few different metastable states. Ironically, the key challenge in discovering the possible nonlocal moves is to remain sufficiently close  to the reference configuration $\textbf{s}^{\star}$ in order to get reliable local information that can eventually help ``jumping out'' or ``escaping'' from the low-energy minimum along a low-energy saddle. This most probably will bring the system close to a different and far away low-energy basin of attraction. 

To discover local information on frozen variables, we keep the reference configuration $\textbf{s}^{\star}$ fixed and introduce a surrogate variable $\textbf{r}$ that initially are set equal to $\textbf{s}^{\star}$. In order to use the surrogate variable $\textbf{r}$ as a probe of the local energy landscape, we evolve it according to the following biased surrogate Hamiltonian $H_\epsilon(\textbf{r}) = H(\textbf{r}) - (\boldsymbol{\epsilon} \circ \textbf{s}^{\star}) \cdot \textbf{r}$, where $\circ$ denotes entrywise product between an inhomogeneous vector $\boldsymbol{\epsilon}$ and reference configuration $\textbf{s}^{\star}$. For large $\lVert \boldsymbol{\epsilon} \rVert $ the surrogate system will stay very close to the reference configuration, while for $\lVert \boldsymbol{\epsilon} \rVert \to0$ the surrogate system becomes an independent replica. It is more convenient to factor out a global scaling parameter $\lambda$ to control the radius of sampling with respect to the reference configuration; that is we rescale $\boldsymbol{\epsilon}$ as $\lambda\boldsymbol{\epsilon}$ where $\boldsymbol{\epsilon}$ is now fixed and only $\lambda$ can vary to control locality of surrogate Hamiltonian. Here, we define $\epsilon_{i} =  |h_i| + \sum_{j}| J_{ij}|$ and initially set $\lambda\gg1$ to ensure that the initial $\lambda\epsilon_i$ for each site is large compared to the energy scale of the site. This inhomogeneous construction of the vector $\boldsymbol{\epsilon}$ guarantees the locality of the surrogate Hamiltonian over certain core variables for problems with highly heterogeneous underlying graph topology: namely variables with many edges and/or very strong couplings; e.g., the hubs in the scale-free networks  with small-world properties \cite{barabasi_emergence_1999}. We can recover the limit that the surrogate system could act as an independent replica for $\lambda\to0$. More explicitly the surrogate local Hamiltonian becomes:
\begin{equation} \label{eq:explicitSurrogateHamiltonian}
H_\epsilon(\textbf{r}) = H(\textbf{r}) - \lambda \sum_i \epsilon_{i} s^{\star}_i r_i
\end{equation}

 Magnetizations $\langle r_i \rangle$ and correlations $\langle r_i r_j \rangle$ of the surrogate Hamiltonian variables depend on ${\epsilon_{i}}$, but we are interested only in the ordering of variables according to some criterion (e.g., decreasing order in magnetization or correlation). Such ordering is preserved in a broad range of $\lambda\boldsymbol{\epsilon}$. This will allow us to introduce a robust mechanism for thresholding correlations to capture the degree of rigidity among variables for a variety of replicas in a fairly large range of temperatures.

\section{Efficient sampling of localized surrogate Hamiltonians via LBP}

\label{sec:LBP}

Belief Propagation (BP) is an iterative message-passing algorithm that solves the self consistency equations obtained within the Bethe approximation, and thus computes approximate marginal probabilities on small sets of variables (e.g.\ magnetizations and correlations) \cite{BP03}. In this context it is similar to tensor-network contraction techniques in quantum many-body physics. Indeed, both techniques can be captured as variants of the Bethe-Peierls approximation in statistical physics \cite{Arad_tensornetwork_BP_2020}. BP is known to be exact only on trees or when graphs have at most one loop \cite{MezardBook}. The convergence and reliability of BP can also be understood in terms of the Bethe approximation which is exact on trees. Generalized Belief Propagation (GBP), inspired from the Kikuchi cluster variational approximation to the Gibbs free energy, can be efficiently extended to situations where there are many frustrated loops, but all such loops need to be local \cite{BP03}. Unfortunately, the complexity of the GBP algorithm grows exponentially with the length scale of the frustrated loops. In practice, however, one can apply BP to loopy graphs, namely Loopy Belief Propagation (LBP), which can return highly accurate local marginals in problems where connected correlations decay fast enough along the interacting graph. This actually corresponds to models having a single pure state \cite{MezardBook}.  

In this work, we show that by properly rescaling the inhomogeneous vector $\lambda\boldsymbol{\epsilon}$ as a localizing penalty term in the Hamiltonian, we can control the surrogate problem to be sampled from the pure state or the basin of attraction that $\textbf{s}^{\star}$ belongs to. Thus, we can safely use LBP for our surrogate Hamiltonian given by Eq. \ref{eq:explicitSurrogateHamiltonian}. For energy functions with pairwise interactions the measure to be sampled is proportional to
\begin{equation} \label{eq:BoltezmannSurrogate}
\exp\left[\beta \sum_{(ij)\in E} J_{ij} r_i r_j + \beta \sum_i (h_i + \lambda \epsilon_{i} s^{\star}_i) r_i\right]
\end{equation}

This is a general Ising model, where the external field has been modified by the presence of the coupling with the reference configuration. The corresponding LBP equations are the following:
\begin{align}
h_{i\to j} &= h_i + \lambda \epsilon_{i} s^{\star}_i + \sum_{k \in \partial i \setminus j} u_{k\to i}\\
u_{i\to j} &= \beta^{-1} \text{arctanh}[\tanh(\beta J_{ij}) \tanh(\beta h_{i\to j})]
\end{align}
where $\partial i = \{j:(ij)\in E\}$ is the set of neighbors of $i$.
These are $2|E|$ equations in the so-called cavity fields and can be solved e.g.\ iteratively. From the solution of the above equations one can obtain the magnetizations as:
\begin{equation}
\langle r_i \rangle = \tanh\bigg[\beta \Big(h_i + \lambda \epsilon_{i} s^{\star}_i + \sum_{j \in \partial i} u_{j\to i}\Big)\bigg]
\end{equation}
and correlations between nearest neighbors, i.e.\ for $(ij)\in E$:
\begin{equation}
 \langle r_i r_j \rangle = \frac{\tanh(\beta J_{ij}) + \tanh(\beta h_{i\to j}) \tanh(\beta(h_{j\to i})}{1+\tanh(\beta J_{ij})\tanh(\beta h_{i\to j}) \tanh(\beta(h_{j\to i})}\;.   
\end{equation}
It is known that a much better estimate of correlations can be achieved via linear response; however, this requires a slower algorithm than LBP.

In order to estimate $\langle r_i \rangle$ and $\langle r_i r_j \rangle$ for many different values of $\lambda \boldsymbol{\epsilon}$, we perform LBP in an adiabatic fashion  by starting from a large $\lambda$ and initialize the LBP messages as $h_{i\to j}=\lambda \epsilon_{i} s^{\star}_i$ and $u_{i\to j}=J_{ij}s^{\star}_i$. After estimation of magnetizations and correlations at each step, the value of $\lambda$ is gradually decreased, but we do not reinitialize the LBP messages: indeed changing $\lambda$ by a small amount leads to small changes in solution, and thus we can converge quickly if we start from the previous solution to LBP which is obtained in the previous step. 

When $\lambda$ becomes too small the surrogate Hamiltonian will start sampling from configurations that are outside the pure state that $\textbf{s}^{\star}$ belongs to. This may lead to either very small values of the overlap with the reference configuration
\begin{equation}
\label{Eq.p_overlap}
p = \frac1N \sum_i s^{\star}_i \langle r_i \rangle,    
\end{equation} 
or lack of convergence of the iterative method to solve the LBP equations. In the latter case, we then use the information collected in the previous iteration.

A possible criterion to understand the range of values of $\lambda$ leading to a sampling within the pure state -- that $\textbf{s}^{\star}$ belongs to -- involve the comparison of the overlap $p$ with the self overlap
\begin{equation}
 q_1 = \frac1N \sum_i \langle r_i \rangle^2   
\end{equation}
Indeed if both $\textbf{s}^{\star}$ and $\boldsymbol r$ are typical configurations of the same state the equality $p=q_1$ holds. In App. \ref{App:localized_replica}, we provide an alternative sampling techniques by cloning Monte Carlo replicas over localized surrogate Hamiltonians. However, this MC-based  approach is not as efficient or as reliable as LBP, since it does not guarantee a linear scaling with input size nor provide any signal if we have left the basin of attraction, which is characterized by $\textbf{s}^{\star}$, in an uncontrolled way.

\section{Efficient sampling of  k-local surrogate Hamiltonians}

\label{sec:LBPhigher}

Before constructing the surrogate backbones and nonlocal moves using the knowledge of LBP, we first consider a generalization of problem classes from 2-local to $k$-local Hamiltonians. In this context, the key ingredients of our algorithm, such as the construction of localized surrogate Hamiltonians and LBP evaluations described in the previous sections, require generalization to higher interacting systems with $k>2$.  These generalizations are important from both fundamental and practical perspectives. They provide us the flexibility of choosing hard benchmark problem instances. Many industrial Max-SAT problems, including those in international SAT and Max-SAT competitions, usually involve clauses with $k \geqslant 3$ variables. Random k-SAT problems near the computational phase transition exhibit average-case hardness involving a first order phase transition for $k \geqslant 4$. The k-local formalism also allows us to implement both the replica-exchange MC and LBP subroutines directly on the CNF formulation, as we show in Sec. \ref{sec:numerics} and App.~\ref{App:LBP_CNF}. This considerably reduces the computational overhead of mapping or embedding the problems to 2-local Ising models, so the algorithm can be implemented much more efficiently and be numerically benchmarked for significantly larger problem sizes involving $O(10^4)$ or more variables. Moreover, these generalizations could be used to abstract-out the advanced spin-glass physics. 

In the first step, we change the representation of the generalized spin-glass systems to be modelled as factor graphs. 
We then generalize our LBP calculations on the surrogate k-local Hamiltonians, including calculations of k-local correlation functions to be able to grow clusters of rigid or frozen variables. 

The factor graph is a bipartite graph where edges connect factor nodes in the $F$ set with variable nodes in the $V$ set. Let us write the generalized Ising model over a factor graph as:
\begin{equation}
H(\boldsymbol r) = -\sum_{a\in F} J_a \prod_{i \in \partial a} r_i - \sum_{i \in V} h_i r_i,
\end{equation}
where $V$ is the set of all vertices, each one representing a single Ising variable; and $F$ is the set of all factor nodes, each one representing a particular multi-spin interaction. We mostly adopt the notations for factor graphs that are consistent with Ref. \cite{MezardBook}. In this notation, $\partial a$ is the set of variables entering in the $a$-th interaction. For pairwise interactions $|\partial a|=2$, while for $k$-spin interactions $|\partial a|=k$. Sometimes, we may use a shorthand notation for the external field, $h_i$, but we remind the reader that whenever we are dealing with the surrogate Hamiltonians, one has to substitute $h_i \to h_i + \lambda \epsilon_{i} s_i^\star$.

 LBP on factor graphs requires keeping track of two distinct types of messages, those from a vertex $i$ to a factor node $a$, $h_{i \to a}$, and those messages from a factor node $a$ to vertex $i$, $u_{a \to i}$. These messages satisfy the following equations:

\begin{align}
h_{i \to a} &= h_i + \epsilon s_i^\star + \sum_{b \in \partial i \setminus a} u_{b \to i},\\   
u_{a \to i} &= \beta^{-1} \text{arctanh}\left[
\tanh(\beta J_a) \prod_{j \in \partial a \setminus i} \tanh(\beta h_{j \to a})
\right],   \nonumber
\end{align}
where $\partial i$ is the set of factor nodes connected to vertex $i$ and $\partial a$ is the set of variable nodes connected to $a$. At convergence the LBP messages can be used to infer local marginals as follows
\begin{equation} \label{Eq:firstmarginals_k-local}
  \langle r_i \rangle = \tanh\bigg[\beta \Big(h_i + \sum_{a \in \partial i} u_{a\to i}\Big)\bigg]  
\end{equation}

Pairwise correlations bring almost no information in high-order interacting models.
The lowest order non trivial correlation is the following
\begin{equation} \label{Eq:highermarginals_k-local}
  \langle \prod_{i \in \partial a} r_i \rangle = \frac{\tanh(\beta J_a) + \prod_{i \in \partial a} \tanh(\beta h_{i\to a})}{1+\tanh(\beta J_a)\prod_{i \in \partial a} \tanh(\beta h_{i\to a})},  \end{equation}
where the first and higher-order marginals can be used to discover the backbones by imposing a threshold cutoff based on certain general criteria as we will describe in Sec. \ref{sec:backbones}. A generalization of our k-local algorithm for the general factor graph on CNF is presented in the App.~\ref{App:LBP_CNF}.

\section{Generating backbones of rigid variables}

\label{sec:backbones}

Here, we outline our main algorithms for growing clusters of connected variables based on LBP sampling. In App.~\ref{App:alternate_backbones}, we outline two alternative methods for creating disconnected clusters that are using simple thresholding of the 2-point correlation functions, and illustrate the basic concepts, but they are not very effective in practice. The main method that we employ in our simulations and benchmarking has a direct physical interpretation. In this method, we grow connected clusters of correlated spins that are forming the backbones of surrogate Hamiltonians and can be understood as droplet-like excitations of the original spin-glass problem. 

In all of our cluster growing algorithms, we first strongly enforce the locality of surrogate Hamiltonians by initially pining each to the basin of attraction characterized by $\textbf{s}^{\star}$ re-scaled by an inhomogeneous vector, $\lambda\boldsymbol{\epsilon}$ with large enough $\lambda$.
Each entry $\epsilon_i$ in the epsilon vectors is set to $\epsilon_i = (|h_i| + \sum |J_{ij}|)$ to ensure that the initial epsilon for each site is large compared to the energy scale of the site. These inhomogeneous $\epsilon$ vector guarantee the locality of surrogate Hamiltonian over the key variables. These heavyweight variables likely belong to the unknown backbone of the problem, but that is not always the case within each pure state of a given replica. Next, we calculate the initial LBP messages $h_{i\to j}$ and $u_{i\to j}$ by doing one iteration over the LBP equations, within the large $\lambda$ limit, of $h_{i\to j} = \lambda \epsilon_i s^{\star}_{i}$ and $u_{i\to j} = J_{ij} s^{\star}_{i}$. We then reduce $\lambda$ incrementally and update the LBP messages accordingly. The criteria for stopping the LBP and how to grow the cluster of rigid variables vary among various strategies for growing clusters.

In our main strategy in this work, we grow a set of connected clusters that provide a direct physical interpretation as a generalization of the droplet excitations which are traditionally studied in the low-dimensional spin-glass systems and recently being characterized by approximate tensor-network contractions for quasi-2D spin glass systems with local fields \cite{rams_tensors_2021, mohseni_diversity_2021}. Historically, the droplet picture for excitations was first introduced in the context the Edwards-Anderson model of spin glasses by D. Fisher and D. Huse \cite{fisher_equilibrium_1988}. In simple terms, droplets are the cheapest spin cluster measured in terms of excitation energy. In principle, if one could efficiently evaluate partition functions exactly, one could use such enormous computational power to create droplets following these steps: one could first evaluate the ground state according to free boundary conditions which yields a reference spin configuration. Then one would fix some boundary spins and flip a random (central) spin, and calculate a new ground state accordingly. Thus, the droplets could be fully characterized by finding the orientation of the spins in the new ground state relative to the reference spin configuration. In finite temperatures, droplets are a collection of highly correlated clusters of variables that are highly likely to be separated from the rest of variables by domain walls or topological defects. These droplets could have compact or fractal boundaries organizing themselves into geometries with embedded hierarchy; sometimes resembling sponge-like structures. Recently, we have used strong-disorder renormalization group and approximate tensor-network contraction techniques to estimate the boundary of droplets for quasi-1D and quasi-2D spin-glasses with local fields \cite{mohseni_engineering_2018,mohseni_diversity_2021}. That information was used as a preprocessing step to develop inhomgeneous quantum annealing algorithms for low-dimensional Ising Hamiltonians with significant speedup. Our work here generalizes such works to higher dimensional systems by dynamically estimating such droplet boundaries during the runtime of our algorithm.    

Algorithmically, in the context of this work, a droplet is basically a large cluster of rigid variables that would not flip via local moves in polynomial time unless the temperature is increased. When local and homogeneous MCMC algorithms are employed, the structure of low-energy states can manifest itself as local fluctuations and adjustments/relaxations over a power-law distribution of droplet sizes. The relaxation time for each droplet grows exponentially with size of the droplet and could be understood as one important mechanism behind extremely long aging of spin-glass systems\footnote{It is fair to say that energetic barriers are not the only bottleneck to relaxation in frustrated models, as entropic barriers can play an important role as well \cite{bellitti_entropic_2021}}. Here we employ LBP on the surrogate Hamiltonian to efficiently estimate the boundary of such droplets on the original problem. This could provide a significant computational speedup, since we can control collective spin updates over such droplets causing significant variations in Hamming distance in configuration space, while keeping the energy of the overall systems fairly constant within a target approximation ratio. This nonlocal mechanism for state transitions in a low energy mini-band can be seen as a classical analogue to quantum many-body delocalization algorithms introduced recently \cite{Smelyanskiy_MBDL_2020}, although they rely on fundamentally different many-body effects.

Here, we first find the smallest possible global $\boldsymbol{\epsilon}$, characterized by the scaling factor $\lambda$, in which LBP iterations still converge within some desired precision, and calculate the single and higher-order marginals according to Eqs.~(\ref{Eq:firstmarginals_k-local}) and (\ref{Eq:highermarginals_k-local}). We then define effective interactions/couplings for 2-local and k-local  Hamiltonians as:  
\begin{equation}
\tilde{J_{ij}} =  \beta^{-1} \arctanh{(\langle r_i r_j \rangle)},
\end{equation}
and
\begin{equation}
\tilde{J_a} =  \beta^{-1} \arctanh{(\langle \prod_{i \in \partial a} r_i \rangle)},
\end{equation}
where $\tilde{J_{ij}}$ and $\tilde{J_a}$ denote the effective coupling and the effective factor node that corresponds to interactions $J_{ij}$ and factor node $J_{a}$ in the original Hamiltonians respectively, and $\langle r_i r_j \rangle$ and $\langle \prod_{i \in \partial a} r_i \rangle$ are two or higher order correlations calculated with LBP. These effective couplings are the key variables that are used for quantifying the rigidity of variables and will be compared against the correlation thresholds for growing clusters. 

In order to grow connected clusters/droplets, we set up two different correlation thresholds that are typically a few percent apart from each other. The first one which we call the \textit{seed correlation threshold} is used to find the seeds of the surrogate backbones. The second one, which we call the \textit{correlation threshold cutoff}, determines the size of such clusters. Specifically all variables with effective couplings larger than the \textit{seed correlation threshold} are selected as a nuclei or seeds to form a cluster. From each seed a connected cluster is grown by adding all neighboring spins connected to a seed spin with effective couplings above \textit{correlation threshold cutoff}. The next spins to be added are those spins outside the cluster that are connected to one or more spins inside the cluster and their effective couplings or marginals are above the threshold cutoff. This is repeated until the correlation between new spins outside the cluster and spins inside the cluster drop below the correlation threshold cutoff and no more spins can be added to the cluster. Then, we move to the next seed and grow it to maximum size such that all effective interactions within the droplet are above the correlation cutoff. We repeat this procedure until there is no more large effective couplings which can qualify as a seed for a new droplet. For alternative strategies to grow disconnected clusters see App.~\ref{App:alternate_backbones}.

\section{Correlation threshold cutoff}

\label{sec:thresholding}

The correlation threshold cutoff is the key parameter in our algorithm which significantly impacts the size and shape of surrogate backbones and consequently the efficacy of the cluster updates or nonlocal moves. There are several important aspects of the correlation threshold that we have examined: (i) We have empirically verified, over several different problem classes, that there exists an acceptable value of the correlation threshold cutoff such that the surrogate Hamiltonian backbones can have meaningful large sizes without percolating (e.g., between $N/20$ to $N/2$); (ii) we have found that the value of the correlation threshold is robust over a wide range of values; i.e., emerging clusters do not percolate suddenly from very small clusters to very large ones in a very small range of values for the correlation threshold cutoff, see App. \ref{App:other_problems}; (iii) we have observed that, within the acceptable range of correlation thresholds, we could  grow backbones that lead to useful nonlocal cluster moves to accelerate the sampling for a single MCMC replica, see  App. \ref{App:other_problems}.   

In our numerics we have adopted two alternative approaches to tune the optimal value(s) of the correlation threshold within an  acceptable range. In the first approach, we use a machine learning technique to train a black-box hyperparameter optimizer. This tool, which is publicly offered by Google cloud platform, known as  \lq\lq Vizier \rq\rq which predominantly employs a Bayesian learning toolbox, such as Gaussian processes \cite{google_vizier}. This approach is more effective in finding the optimal value of the correlation threshold on the runtime in an instance-wise fashion as we demonstrate for solving hard instances of QAP in Sec. \ref{sec:numerics}. In the second approach, which was adopted for solving hard random 4-SAT instances near phase transition, we develop a quantum-inspired approach and \lq\lq adiabatically\rq\rq\ anneal the values of the correlation thresholds from relatively low values (with cluster sizes of $O(N)$) to very conservative values near unity (with the maximum size of clusters in single digits). For more details on this variant of NMC algorithm see App. \ref{App:adiabaticNMC}. We will discuss how the NMC algorithm will arrive at its nonequilibrium steady state in the section.

\section{Nonequilibrium inhomogeneous sampling  over subgraphs}
\label{sec:balance}

In this section, we describe our inhomogeneous MCMC algorithm and discuss its nonequilibrium steady states. We first focus on the simplest possible scenario which is a single replica in a single round of APT. Using our estimation about the backbone's boundary in a given basin of attraction, we construct two different Markov chains at two different high and low temperatures for variables inside and outside of the backbone respectively. In our numerical study, the temperature of backbone variables is typically elevated from the rest of variables by a factor of 2 to 10, although extremely high temperatures could become beneficial for certain backbones. We then iterate between these two Markov chains with a relatively high frequency.
Since any finite Markov chain whose transition probabilities do not have an explicit time dependence admits  a stationary distribution, both chains in each iteration are able to arrive at a new stationary state over their corresponding subgraphs, which asymptotically lead to a steady state for the combined system. Next we argue how our algorithm can robustly sample from the relevant low energy manifold of the problem, despite not satisfying a global detailed balance condition. 

Let's assume that we can perfectly identify all the frozen variables; that is, we can grow the correct backbone for each localized surrogate Hamiltonian after thresholding the marginals. Given that assumption we argue that each of the two inhomogeneous MCMC for two induced subgraphs (backbone and non-backbone variables), for a given replica in a given cycle, can efficiently sample from their corresponding low energy states. We note that the inhomogeneous (high-T) MCMC by construction obeys detailed balance over the subgraph defined by the backbone and thus has a stationary state. Thus in principle we will be able to sample in a reversible fashion from all the basin of attractions corresponding to the low energy configurations of that backbone. We also note that for each basin of attraction we are sampling over the complement graph with another MCMC (over those variables not in the backbone). By construction, this second (low-T) MCMC also satisfies detailed balance, implying efficient sampling from low energy states corresponding to a single basin of attraction. Therefore, by iterating these two MCMC, over all the replicas residing in various base temperatures in various APT rounds, we are reversibly sampling all their basins of attraction and their low energy states asymptotically. Therefore, we are effectively sampling all the relevant low energy states that can make major contributions in evaluating the partition function without strictly satisfying detailed balance globally, assuming we have full characterization of the backbones.

Given the fact that we cannot directly verify the ground truth for the surrogate backbones, in practice we invoke unlearning phases in which we apply standard (local and homogeneous) MCMC frequently (after each application of the inhomogeneous MCMC) to mitigate the inductive bias in our model for the backbones and thus smooth out our sampling mechanism. Consequently, we can asymptotically arrive at a global steady state which essentially captures the important low energy properties of the original problem. This leads to a robust performance of our algorithm as evident by our numerical simulations. In fact, NMC exhibits significantly less fluctuations for the best seen states across several repetitions compared to standard APT, while capturing significantly higher diversity in exploring the configuration space within one repetition. This is numerically verified using the whitening procedure (see Sec. \ref{sec:whitening}).

\section{Numerical simulations}
\label{sec:numerics}

We have applied our framework to several classes of discrete optimization problems. Here, we mainly focus on reporting numerical simulations on two important and well-studied classes of hard discrete optimization problems: Random K-SAT  \cite{moore_nature_2011}, and Quadratic Assignment Problems (QAP) \cite{burkard_QAP_1998}. In the first class, we studied random 4-SAT problems with 5000 variables deep into the rigidity regime. In the second class we benchmarked our algorithm on random QAP instances, with sizes ranging from 256 to 1600 binary variables \cite{drugan_generating_2015}, as well as some industrial instances from the QAPLIB public library \cite{burkard_qaplib_1997}. We also applied NMC algorithm to other combinatorial optimization problems with various dimensionality and structure. We observed that our main algorithmic subroutines for finding frozen variables, discovering computationally relevant surrogate backbones, and building useful inhomogeneous MCMC perform reliably across different problem classes. In App. \ref{App:other_problems}, we provide a few examples from structured instances with local fields on the Chimera graph, based on the architecture of the D-Wave quantum processors with quasi-2D geometry, and other structured instances from weighted Max-Cut problems. 

For various problems we have generally employed three main performance metrics: (i) quality of the solutions (e.g., number of violations, approximation ratio, or the residual energy) obtained in a given time, (ii) success rate per run/repetition, and (iii) time to arrive at an approximate solution with a desired quality. For the random 4-SAT problems we additionally investigated the existence and size of frozen clusters in the best found solutions using a whitening procedure \cite{parisi_local_2005}. We compared the performance of NMC against various generic stochastic solvers such as APT and WalkSAT, and several best known deterministic SAT solvers, based on Conflict Driven Clause Learning (CDCL) \cite{biere_handbook_2021} or core-guided Max-SAT solvers, such as  MiniSat \cite{Goos_extensible_2004}, RC2 \cite{rc2_2019}, EvalMaxSat, and state-of-the-art specialized message-passing solvers such as Survey Propagation and BSP \cite{marino_backtracking_2016} (for a summary of SP and BSP algorithms see App. \ref{App:SP-BSP}). The benchmarking was performed on Google's distributed computing platform~\cite{borg_2015} and the automatic hyper parameter optimization was performed with Vizier~\cite{google_vizier}. 

\subsection{4-SAT Problems Near Computational Phase Transitions}

The Boolean Satisfiability (SAT) \cite{pulina2020theory, yolcu2019learning} is the problem of determining if there is an assignment that satisfies a given Boolean formula. A Boolean formula is any operation made with Boolean variables, where each variable can take the value $TRUE$ or $FALSE$, or $\{1, 0\}$ respectively. For example, a CNF (conjunctive normal form) \cite{chang2021boolean} formula is a conjunction of one or more clauses, where a clause is a disjunction of literals. When there are exactly $k$ literals for each clause in a CNF formula, the problem is named $k$-SAT. A CNF formula is satisfiable if and only if the Boolean variables' configuration satisfies all the clauses simultaneously. The $k$-SAT problem for $k\geq 3$ is central in combinatorial optimization: it was among the first problems that were characterized  as NP-complete \cite{cook1971complexity, garey1979computers}.

The maximization problem associated with $k$-SAT is called MAX-E-$k$-SAT. In this case, a solver tries to satisfy the maximum number of clauses given a CNF formula of the $k$-SAT problem. The MAX-SAT is of considerable interest not only from the theoretical side but also for applications. For instance, software and hardware verification problems, automated resonating, and several open problems in artificial intelligence such as training and inference in graphical models can be expressed in the form of satisfiability or some of its variants. From the theoretical point of view, the MAX-SAT problem is studied to give optimal inapproximability results. For the $k$-SAT problem, the most important work given for inapproximability was due by H{\aa}stad in 1997 \cite{haastad2001some}. He proved optimal inapproximability results, up to an arbitrary $\epsilon>0$, for MAX-E-$k$-SAT with $k \geq 3$. The approximation algorithms do not tell us how well we might be able to do, instead they will tell us how hard is to satisfy the sufficient condition for the worst-case, i.e., how badly we might perform. Here, we want to explore how well we can approximate k-SAT for smallest value of $k$, that is $k=4$, such that even median instances are hard to solve for sufficiently dense clauses to variable ratio near the computational phase transition \cite{montanari-clusters-2008}.

In order to benchmark our algorithms on random 4-SAT instances, we developed an adaptive PT that works directly on the CNF formulation with arbitrary k-local clauses. We obtained 10x wall-clock time speedup by running PT directly on CNF instead of an Ising formulation of the problem. We then generalized the LBP algorithm for general factor graphs representing a k-local CNF formulation. Here, we provide a summary of the main results; for more details see App.~\ref{App:LBP_CNF}. Using the notation introduced in Sec. \ref{sec:LBPhigher}, the LBP equations over a CNF Boolean formula, for messages $h_{i \to a}$ from variable/literal $i$ to clause $a$ and $u_{a \to i}$  from clause $a$ to variable/literal $i$, can be written as:

\begin{equation}
h_{i\to a} = \lambda \epsilon_i s_i^{\star} + \sum_{b \in \partial i ^+ \setminus a} u_{b \to i}- \sum_{b \in \partial i ^- \setminus a} u_{b \to i},
\end{equation}
where $b \in \partial i ^+ \setminus a$ denotes the set of clauses in $\partial i$ agreeing with clause $a$ on what values $i$ should take. Similarly, $b \in \partial i ^- \setminus a$ denotes the set of clauses in $\partial i$ disagreeing with clause $a$ on what values $i$ should take. The messages from clause to variables, $u_{a \to i}$, satisfy: 
\begin{equation}
u_{a \to i} = -\frac12 \ln\left[1 - (1-e^{-2\beta})\prod_{j \in \partial a \setminus i} \frac{1-\tanh h_{j\to a}}{2}\right]\nonumber\;.
\end{equation}

Using the above relations, the local magnetization or polarization of variables becomes:
\begin{equation}
\langle r_i \rangle =  \tanh\left(\lambda \epsilon_i s_i^{\star} + \sum_{a \in \partial i ^+} u_{a \to i}- \sum_{a \in \partial i ^-} u_{a \to i}\right)\;.
\end{equation}

%%%%%%%%%%%%%%%%%%%%%%%%%%%%%%%%%%%%%%%%%%%%%%%%%%%%%%%%%%%%%%%%%%%%%%%%%
\begin{figure}[t]
    \centering.
    \includegraphics[width=1.\columnwidth]{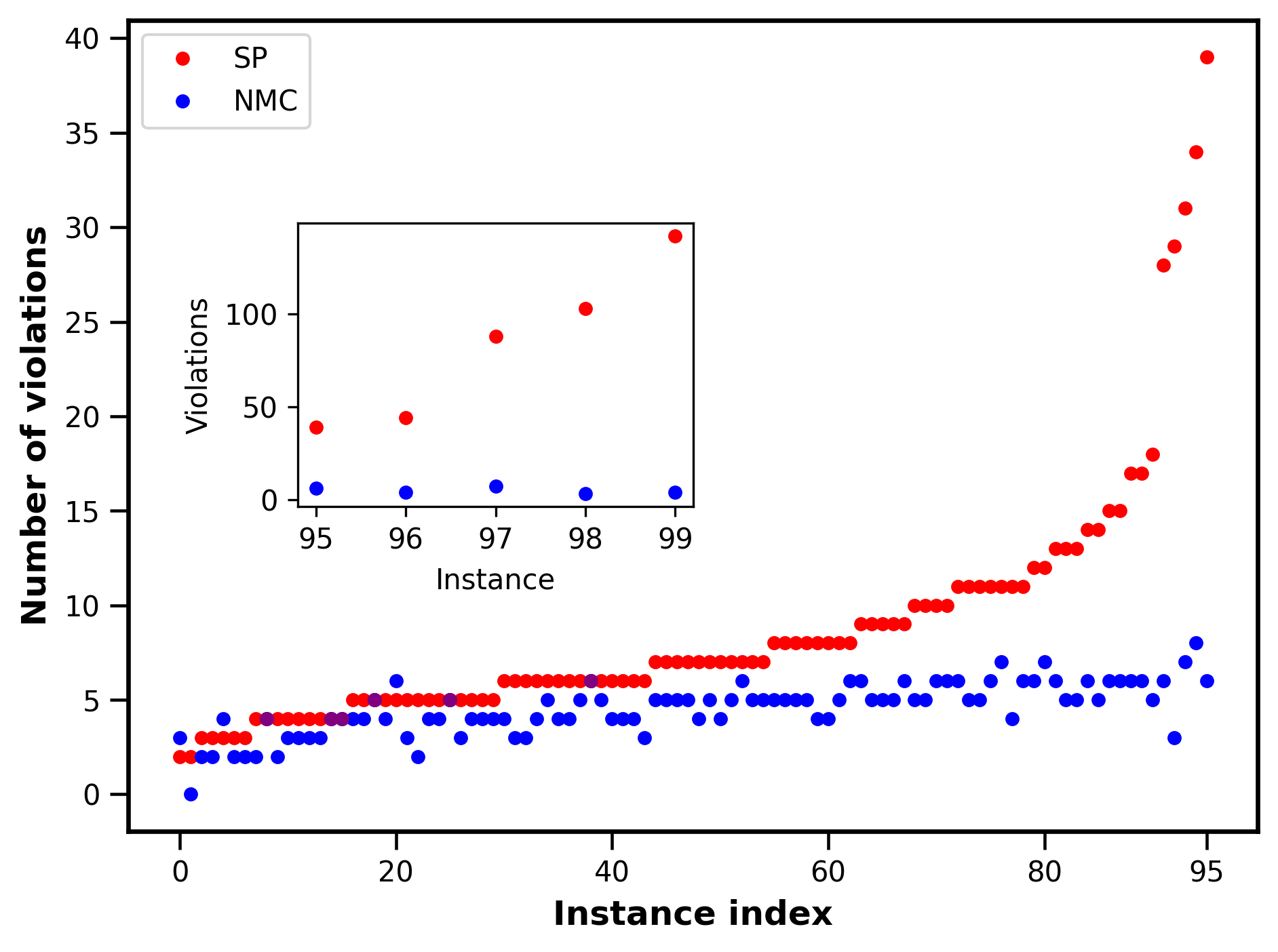}
    \caption{Instance-wise comparison for the best known deterministic algorithm, Survey Propagation (SP) \cite{mezard_analytic_2002}, against Nonequilibrium Monte Carlo (NMC) for a single run on 100 random 4-SAT instances with 5000 variables and $\alpha=9.884$ near a computational phase transition. We note that SP solve only 30\% of the instances, within an approximation ratio of $10^{-4}$, which is equivalent to less or equal 5 violations. NMC solves 75\% of such instances in the same approximation ratio with only four repetitions at $10^{9}$ total sweeps. For about 10\% of all instances NMC obtains at least one order of magnitude improvement in the quality of solutions against SP. It is worth noticing that NMC exhibiting a high degree of robustness, as the worst case is very similar to the median one in sharp contrast to the SP algorithm with significant performance dispersion across various instances.}
    \label{fig:SP_BSP_instanceswise}
\end{figure}
%%%%%%%%%%%%%%%%%%%%%%%%%%%%%%%%%%%%%%%%%%%%%%%%5%%%%%%%%%%%%%%%%%%%%%%%

The high-order correlation function is obtained as:
\begin{equation}
\begin{split}
\langle \prod_{i \in \partial a} r_i \rangle = \frac{-(1-e^{-2\beta})\prod_{i \in \partial a} J_i^a \frac{1-\tanh h_{i\to a}}{2}} {1-(1-e^{-2\beta}) \prod_{i \in \partial a} \frac{1-\tanh h_{i\to a}}{2}},
\end{split}
\end{equation}
where $(J^{a}_1, J^{a}_2, ..., J^{a}_k)\in \{-1,+1\}^k$ are the set of constants that define the constraint represented by clause $a$ involving k variables. 

We generated 100 random 4-SAT instances each containing 5000 variables with a clauses to variables ratio of $\alpha=9.884$. These instances are essentially deep into the rigidity region, with the rigidity threshold of $\alpha_{r}=9.883$ estimated with cavity methods \cite{montanari-clusters-2008,marino_backtracking_2016}. At this value of $\alpha$ for the 4-SAT near the SAT/UNSAT computational phase transition, the instances are believed to be median case NP-hard and exhibit random first order phase transitions, whereas the 3-SAT instances are worst-case hard and undergo a second order phase transition \cite{montanari-clusters-2008}. Thus one expects that generic SAT or Max-SAT solvers, such as APT or CDCL-based algorithms experience an exponential increase in runtime to solve these 4-SAT instances, or approximate with a constant cost, even for typical cases. Indeed, we have also tried several generic solvers including  MiniSat \cite{Goos_extensible_2004}, RC2 \cite{rc2_2019}, and EvalMaxSat, that have been among top performing solvers over previous years of SAT and Max-SAT competitions in many different categories. Neither solvers could generate a meaningful output on any of the instances in several weeks. Thus, these instances exhibit exponential hardness for these classes of deterministic SAT or Max-SAT solvers as expected. In general, SP and BSP are recognized as the current best solvers for this class of problems at sufficiently large sizes \cite{marino_backtracking_2016}.  Thus, here we mainly focus on comparing and contrasting the best approximate solutions and the minimal number of violations computed by NMC, APT, SP, BSP, and WalkSAT algorithms.

%%%%%%%%%%%%%%%%%%%%%%%%%%%%%%%%%%%%%%%%%%%%%%%%%%%%%%%%%%%%%%%%%%%%%%%%%
\begin{figure}[t]
    \centering
    \includegraphics[width=1.\columnwidth]{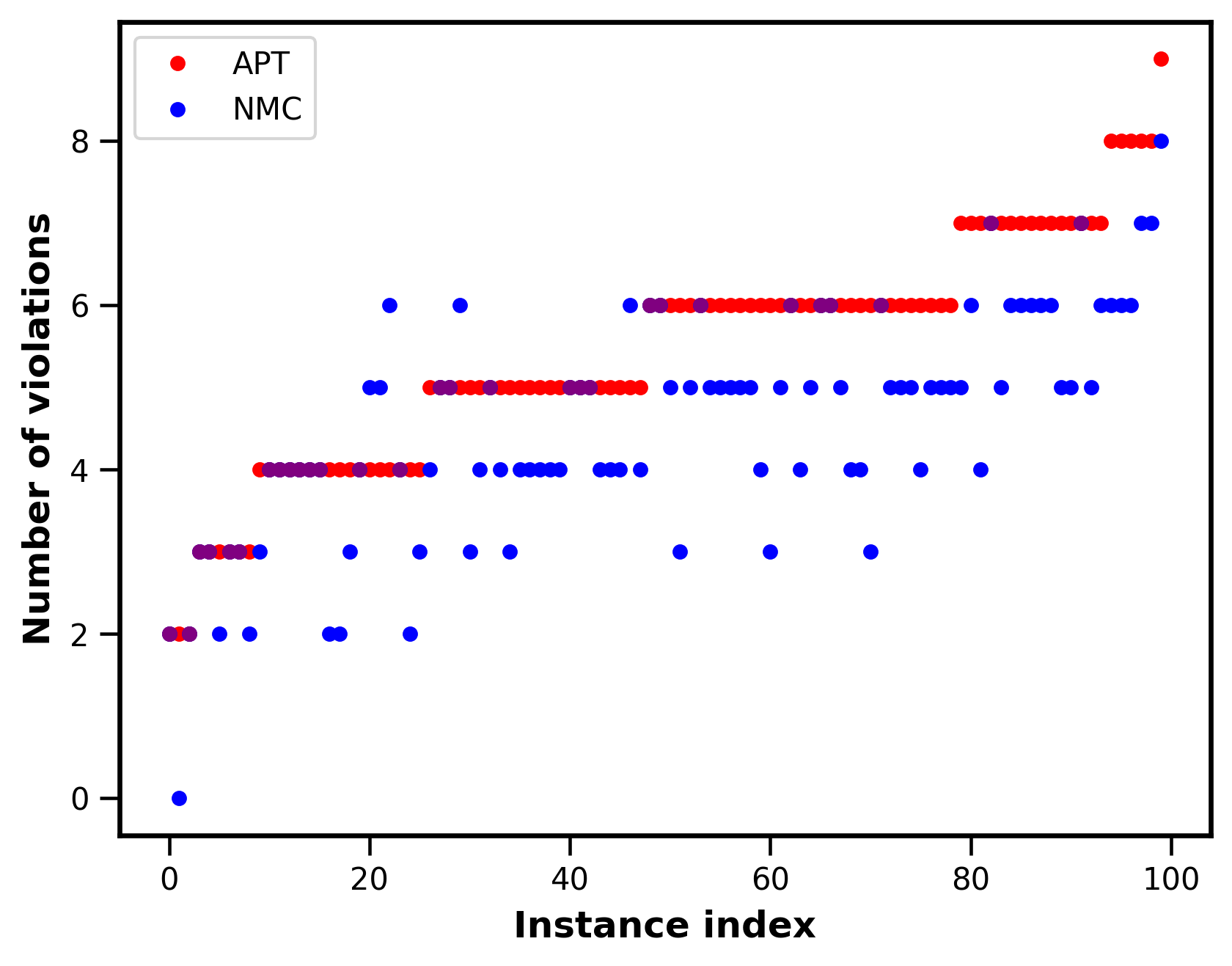}
    \caption{Instance-wise comparison of the best result of four repetitions of APT and NMC for random 4-SAT instances at $10^9$ sweeps. We observe that NMC performs equally or better for 95\% of instances. At this low approximation ratio, resolving each extra  violation could amount to an order of magnitude computational effort. We find that about 70\% of low energy states, obtained by NMC, have frozen clusters, and a majority of those states remain inaccessible with APT even with $O(1000)$ of repetitions (see Sec. \ref{sec:whitening} on the whitening procedure).}
    \label{fig:APT_vs_nonlocal_instance_wise}
\end{figure}
%%%%%%%%%%%%%%%%%%%%%%%%%%%%%%%%%%%%%%%%%%%%%%%%5%%%%%%%%%%%%%%%%%%%%%%%%%

%%%%%%%%%%%%%%%%%%%%%%%%%%%%%%%%%%%%%%%%%%%%%%%%%%%%%%%%%%%%%%%%%%%%%%%%%
\begin{figure}[t]
    \centering
    \includegraphics[width=1.\columnwidth]{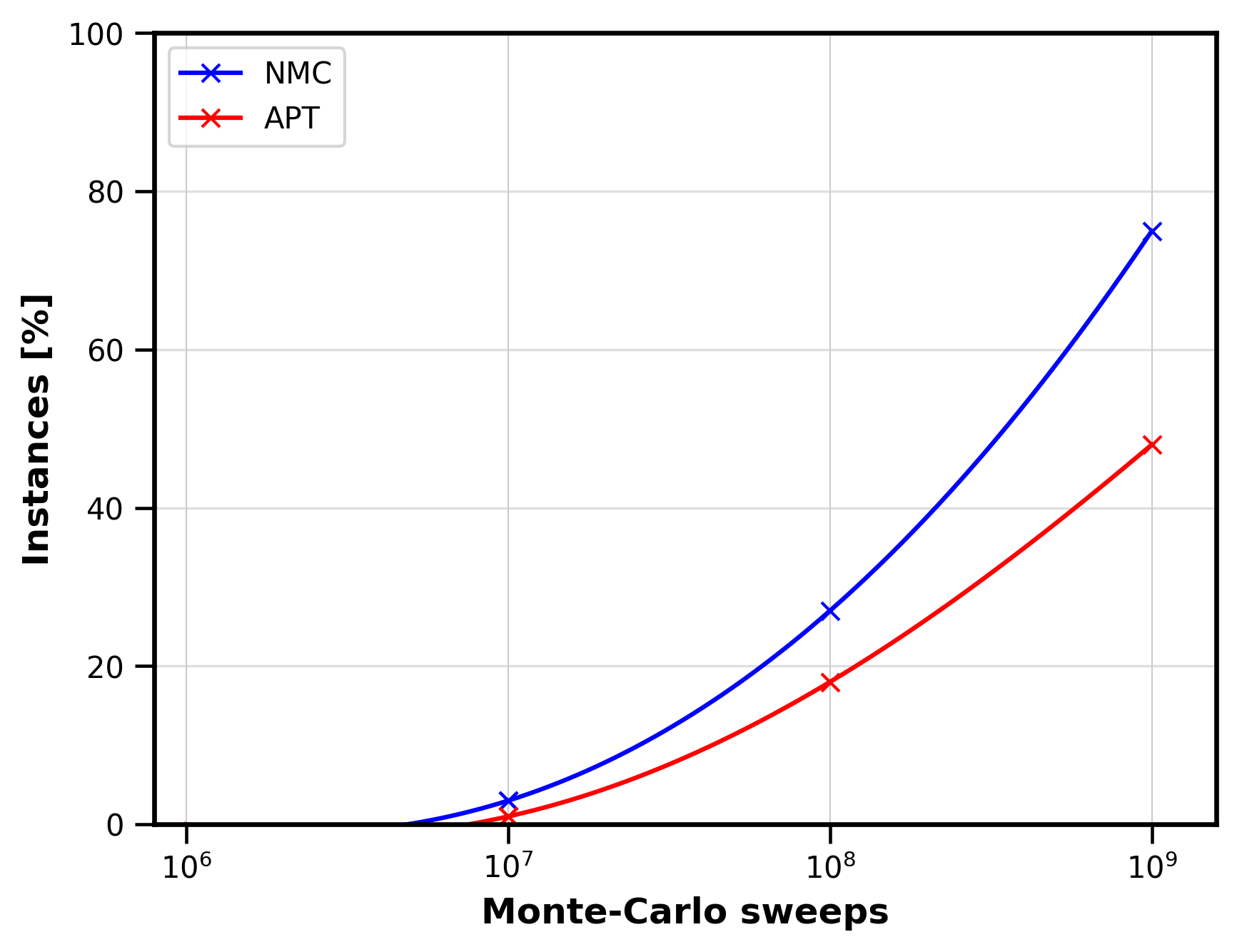}
    \caption{The percentage of random 4-SAT instances solved by APT and NMC within an approximation ratio of $10^{-4}$ as a function of total MC sweeps. We observe that the advantage of nonlocal moves becomes more pronounced with increasing number of MC sweeps for harder instances. In other words, APT will start to saturate at around 50\% of instances at $10^9$ total sweeps while the nonlocal strategy can solve at least 75\% of instances and seems to be far from reaching a plateau of performance. For each solver we had 4 repetitions in each time-scale. The inhomogeneous MCMC runs, in various backbone-induced subgraphs, and the global equilibration (unlearning) MCMC sweeps, are all included in the overall computational effort that is reported for NMC, which is denoted by total number of sweeps.}
    \label{fig:4SATfraction}
\end{figure}
%%%%%%%%%%%%%%%%%%%%%%%%%%%%%%%%%%%%%%%%%%%%%%%%5%%%%%%%%%%%%%%%%%%%%%%%%%
%%

%%%%%%%%%%%%%%%%%%%%%%%%%%%%%%%%%%%%%%%%%%%%%%%%%%%%%%%%%%%%%%%%%%%%%%%%%
\begin{figure}[t]
    \centering
    \includegraphics[width=1.\columnwidth]{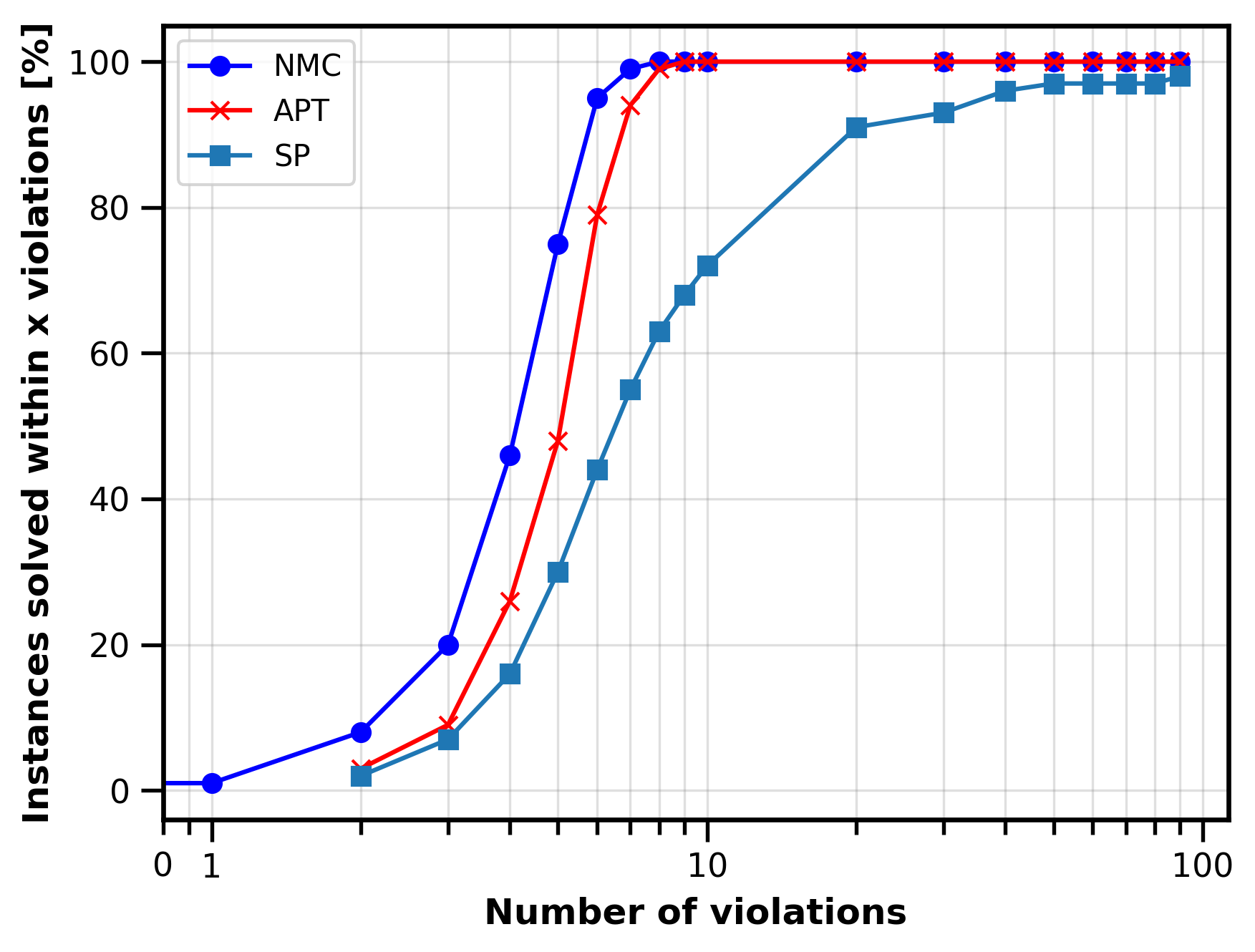}
    \caption{Cumulative percentage of 100 random 4-SAT instances that were approximated by various solvers, Nonequilibrium Monte Carlo (NMC), Adaptive Parallel Tempering (APT), and Survey Propagation (SP), for a given number of violations. Each instance includes 5000 variables with 4-body interactions. The instances are generated near the computational phase transition with a clause to variable ratio of 9.884. NMC solves more than 75\% of all instances within an approximation ratio of $10^{-4}$, that is less or equal to five violations, with only four repetitions at $10^9$ total sweeps compared to 30\% and 45\% for SP and APT respectively.  
    We note that SP obtains about 2x, 10x and 100x more violations than NMC for 75\%, 95\% and 100\% percentile instances respectively.}
    \label{fig:Cumulative}
\end{figure}
%%%%%%%%%%%%%%%%%%%%%%%%%%%%%%%%%%%%%%%%%%%%%%%%5%%%%%%%%%%%%%%%%%%%%%%%

Fig.~\ref{fig:SP_BSP_instanceswise} shows the number of violations, or cost, for each instance using NMC at $10^{9}$ total sweeps in comparison with best solutions found by SP. We note that NMC outperforms SP on 95\% of instances  and could solve 75\% of them within an approximation ratio of $10^{-4}$, which is equal to having 5 or less violations from a total of $49420$ clauses. In contrast SP is able to solve only 30\% of all instances within the same approximation ratio, and no instance to ground state. For the hardest 10\% instances the quality of solutions are improved by NMC algorithm by an order of magnitude. Since SP is a deterministic solver, the hard instances that can not be solved in a target approximation ratio remain inaccessible by this solver irrespective of any arbitrary additional computational time.

In order to be conservative on our estimation of the approximation ratio, we assumed all instances are satisfied at this $\alpha$ which is smaller than the critical SAT/UNSAT value of $\alpha_s=9.931$. However, this is not the case for some of the instances as they are not strictly at the thermodynamic limit. More importantly, the outcome of the NMC algorithm is very robust as the worst case result is close to the median. This is expected for standard MCMC-based algorithms, but in our new algorithm the proposed moves are more nonlocal and keep the system out of equilibrium - a situation where one would have expected more sample to sample fluctuations. On the contrary, there are significant fluctuations in the performance of SP, as well as BSP, across various instances as we will discuss below and in App.~\ref{App:BSPperf}.

An instance-wise comparison of APT and NMC is presented in Fig. \ref{fig:APT_vs_nonlocal_instance_wise}, for best of four repetitions at $10^9$ sweeps, where NMC matches or outperforms APT for 95\% of instances. For random 4-SAT at the rigidity threshold and very low approximation ratio of about $10^{-4}$ resolving every single violations often amounts to an order of magnitude increase in computational resources for local stochastic solvers such as APT. We note that NMC additional subroutines, including LBP runs and inhomogeneous MCMC sweeps, usually add up to a computational overhead of 5\% to 15\% on top of the baseline APT for these instances.   

%With the assumption of $0.3$ spin updates per nanosecond, the $10^9$ total sweeps roughly translate to 6 hours runtime, including the overhead related to running LBP subroutines. However, our actual implementation was slower by 30\% due to some additional computational overheads including latency associated with our large scale distributed systems. 

The fraction of instances that were approximated by APT and NMC as a function of the number of sweeps is shown in Fig \ref{fig:4SATfraction}, where nonlocal moves become significantly more advantageous as we employ more MC sweeps to tackle increasingly harder instances. We observe that a majority of instances remain out of reach for APT as we increase the computational effort by orders of magnitude but the nonlocal strategy keeps finding high quality states for harder instances, presumably penetrating exponentially tall barriers created by large frozen backbones. In order to quantify the underlying cause of slowdown for APT, we use a technique known as whitening procedure (see Sec. \ref{sec:whitening} for more details). Using this approach, we characterize all high quality solutions that are obtained by NMC and observe that 74 instances contain a frozen backbone of size $O(N)$, where a great majority of such frozen clusters are absent for solutions found by APT. For some of these instances, the cluster of solutions with frozen backbone were not observed with the APT algorithm even with up to $O(1000)$ repetitions. We present the cumulative percentage of instances that were approximated with NMC, APT and SP for a given number of violations in Fig \ref{fig:Cumulative}. 

We have also investigated the performance of BSP as the best known specialized stochastic solver for random k-SAT problems. BSP consists of an important stochastic procedure that ideally compiles the original formula into smaller and easier residual formula that can be efficiently handled by a standard WalkSAT solver. BSP employs the information, or beliefs, that become available at the fixed point of standard SP to build this stochastic procedure by applying iterative Survey-Inspired Decimation (SID) or backtracking over subsets of variables with higher marginal probability distributions.
For a short description of SP and BSP algorithm see App. \ref{App:SP-BSP}.

\begin{figure}[t]
    \centering
    \includegraphics[width=\columnwidth]{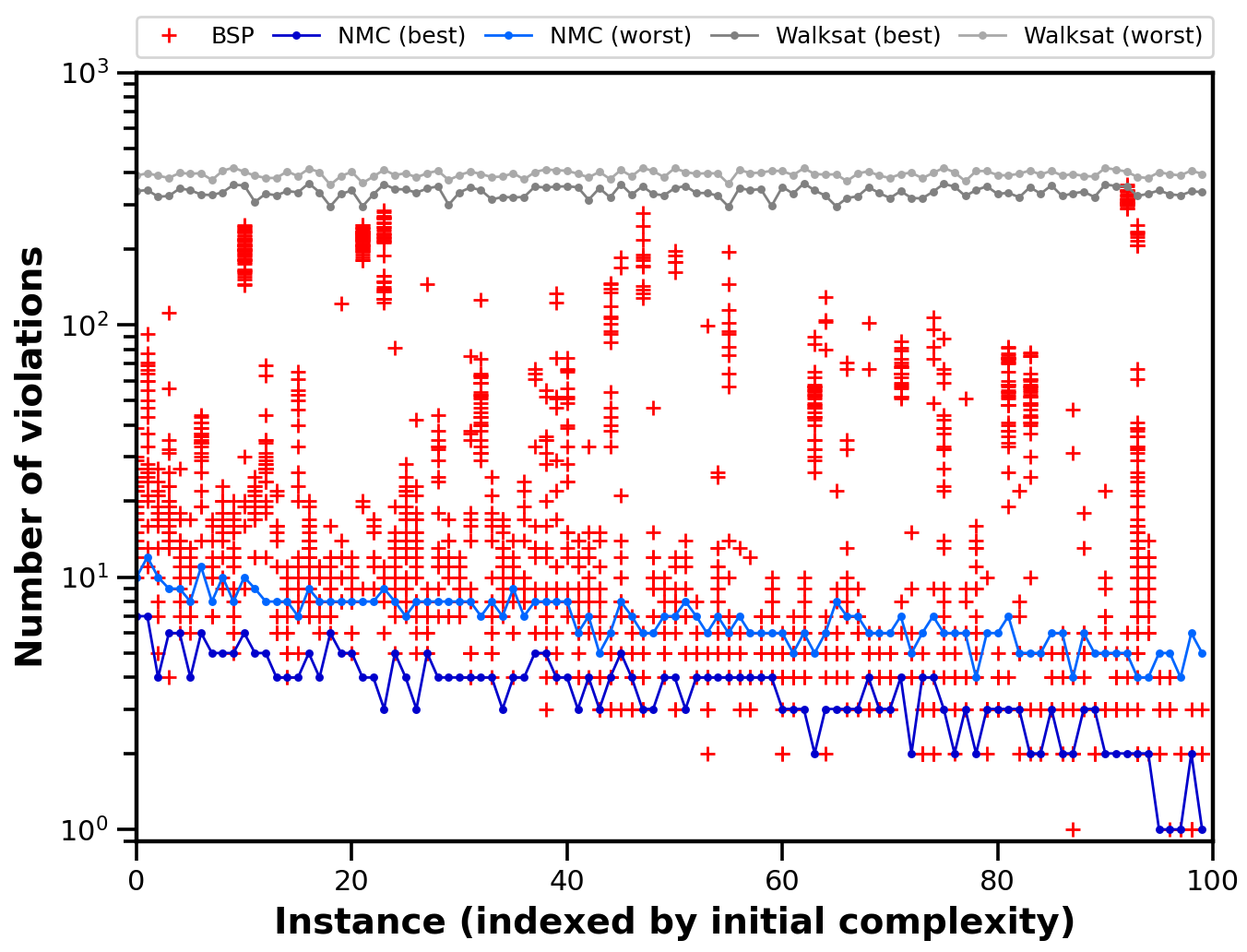}
    \caption{The number of violations obtained by BSP, NMC, and WalkSAT solvers in various repetitions across 100 randoms 4-SAT instances ordered by increasing initial complexity. We see a strong correlation of BSP best runs with this measure. There are significant fluctuations among repetitions for BSP with large degeneracy over the poor solutions leading to very small success probability of obtaining low-energy states for 25\% hardest instances (see App. \ref{App:BSPperf}). NMC also shows strong correlations with this instance-wise hardness measure but with a consistent performance in its worst runs across all initial complexities indicating ability to exploit instances-wise geometrical features in a robust fashion. We note that WalkSAT, as a generic stochastic SAT solver, performs extremely poorly across all instances and all runs, with its best results out of 200 repetitions over 12 hours wall-clock time being two order of magnitude worse in the quality of solutions over the worst NMC runs in 50 repetitions.}
    \label{fig:BSPfulldata}
\end{figure}

The number of violations obtained for up to 50 repetitions of BSP and NMC and 200 repetitions of the standard (pure) WalkSAT algorithm are shown in Fig. \ref{fig:BSPfulldata}. We observe significant performance fluctuations for BSP in various repetitions across all instances. 
The wall-clock time of BSP for each instance was about five hours, for a high backtracking rate of $r=0.999$ \cite{marino_backtracking_2016}, which is a factor two faster than a typical runtime of NMC (between blue curves). The WalkSAT runtime is comparable at 12 hours per instance. A typical run of NMC finds solutions between one to two orders of magnitude better than the best of WalkSAT runs for all instances. We can also see that BSP performs much better than WalkSAT on the best repetitions but their worse runs could become comparable to WalkSAT for about 30\% of instances. The strong fluctuation of BSP is indeed  related to the size of the residual subformulas when WalkSAT is applied, see App.~\ref{App:BSPperf}. Whenever the residual formula is constituted by a small number of clauses, the WalkSAT can return excellent results. However, for a large number of residual clauses, it becomes practically impossible for WalkSAT to find a high quality assignment for the subformula leading to a large number of violations. The latter cases  are indeed computationally as inefficient as a pure WalkSAT run on the original formula. 

Remarkably, the best performance of BSP runs is strongly correlated with initial complexity of each instance (see detailed discussion in App.~\ref{App:BSPperf}). To highlight this feature in Fig.~\ref{fig:BSPfulldata}, we have ordered the instances according to their intial complexity, $\Sigma$, that can be estimated using standard SP on original problem/formula before any decimation. The complexity $\Sigma$ is related to the number of clusters of solutions $\mathcal{N}_{\text{clu}}$ as:
\begin{equation}
\begin{split}
   &\Sigma = \log(\mathcal{N}_{\text{clu}})= \sum_{i=1}^{N} {\Sigma_i} + \sum_{a=1}^{M}(1-|\partial_a|){\Sigma_a};\\ 
   \end{split}
   \label{Eq:complexity1}
\end{equation}
where
\begin{equation}
\begin{split}
&\Sigma_a=\log(1-\prod_{j \in \partial_{a}} \eta_{j \to a});\,\,\,  \Sigma_i=\log(1-\pi^{+}_{i}\pi^{-}_{i});
\end{split}
\label{Eq:complexity2}
\end{equation}
\begin{equation}
\pi_i^{\pm}=1-\prod_{b\in \partial_i^{\pm}}(1-\eta_{b \to i});
\label{Eq:complexity3}
\end{equation}
and $|\partial_a|$ is the length of clause $a$ (initially $|\partial_a|=k$). Here $\eta_{a\to i}$ or $\eta_{i\to a} \in [0, 1]$ are messages or surveys in SP that can be interpreted as the probability that the clause $a$ (variable $i$) sends a message to variable $i$ (or clause $a$) respectively \cite{parisi2003probabilistic, maneva_new_2007}; see App.~\ref{App:SP-BSP} for more details. We note the number of violations for best run of BSP are inversely proportional to instance complexity.  In App.~\ref{App:BSPperf}, we consider this initial complexity as a candidate for instance-wise hardness measure and discuss the origins of the instance dependent large and small fluctuations for BSP and NMC respectively.

Here, it is worth stressing that in the present work we have done a comparison between NMC and BSP running in the playground which is in principle ideal to BSP, that is random $k$-SAT instances.
Indeed the BSP algorithm has derived from the SID algorithm (see App.~\ref{App:SP-BSP}), which is based on the analytical solution to random $k$-SAT problems obtained via the Bethe approximation and the cavity method.
Such an approximation is valid for graphs which are locally tree-like and random graphs have this key feature.
Moving away from random instances we expect BSP, as well as any message-passing algorithm based on the Bethe approximation, to perform much more poorly.
In particular, SAT instances derived from real world application (e.g.\ industrial instance in the SAT competition) are often rich in loopy structures and motifs that make them far from random instances.

%\begin{widetext}
%\end{widetext}

On such non-random instances we expect BSP to face several problems and limitations.
Indeed, the presence of short loops breaks the main assumption underlying the Bethe approximation, where one assumes the probability distribution over the neighbours of any given variable $x$ can be factorized once conditioning on the value of $x$ on itself \cite{MezardBook}.
The breaking of this factorization assumption generates correlations between cavity messages arriving on variable $x$, which in turn have two main effects on the corresponding message-passing algorithm: (i) the iterative solution to the cavity equations may not converge to any fixed point and (ii) even if convergence is achieved, marginal probabilities may be poorly estimated.

The lack of convergence of message-massing algorithms has been clearly measured in the low temperature phase of disordered models, where the effect of frustration becomes particularly strong \cite{parisi2014diluted}. The lack of convergence is enhanced if the model is defined on a regular lattice, or problems with fully connected graphs, due to the presence of many short or intermediate-scale loops \cite{dominguez2011characterizing}. Moreover the presence of loops makes marginals often inaccurate and thus their use (e.g.\ in inference problems) may lead to poor results \cite{ricci2012bethe}.
We expect all these problems to arise when running BSP on non-random SAT instances, or problems with underlying structured scale-free networks, or other high-dimensional problems. For all of those applications, such as QAP that we will study in the next section, SP and BSP will not be competitive with NMC.

\subsection{Quadratic Assignment Problems}

QAP is one of the hardest discrete optimization problems, with variables often residing on a fully connected graph. Thus it serves as a practically relevant use case for our NMC algorithm on a high dimensional problem class \cite{burkard_QAP_1998}. QAP was originally introduced by Koopmans and Beckmann as the problem of allocating a set of  indivisible resources (e.g., economical activities or facilities) to a certain set of locations \cite{koopmans_assignment_1957}. The cost of each particular allocation generally depends on the distance between facilities and their pair-wise flows plus the placement cost of a particular facility at a given location. The problem is finding the optimal assignment of the facilities to the locations that minimize the total cost. QAP is NP-hard and can be formulated as a Quadratic Integer Program, which is a generalization of binary Linear Integer Programming. QAP can also be represented by Quadratic Unconstrained Binary Optimization (QUBO) \cite{burkard_QAP_1998} which becomes equivalent to highly structured fully-connected spin-glass Hamiltonians containing significant disorder and frustration.

We first establish the performance of the APT algorithm on a set of random QAP problem instances as introduced in \cite{drugan_generating_2015}. These instances are designed to be hard for both generic and specialized solvers and, by construction, their optimal solutions are known which greatly facilitates the benchmarking. In Table \ref{table:runs}, we compare the performance of our APT against some well-known generic and dedicated solvers, including Tabu, Glim, Eilm, and GRASP, on 7 random QAP instances with various sizes from 256 to 1600 binary variables \cite{drugan_generating_2015}. The best solutions found by each solver and their non-zero gaps to the optimal solutions were provided in Table 3 of Ref. \cite{drugan_generating_2015}. Remarkably, the optimal solutions for all of these instances were obtained by our APT in only $10^{5}$ MC sweeps, which corresponds to a few seconds wall-clock time. In contrast the gap to optimal solutions varies from 5\% to 80\% for all other solvers across these instances. These results indicate that APT is a very effective generic algorithm to solve this class of random QAP problems, thus invoking nonlocal moves for these instances was unnecessary. 

\begin{table}[]
\begin{tabular}{|c|c|c|c|c|c|c|}
\cline{3-7}
\multicolumn{2}{c}{} & \multicolumn{5}{|c|}{\textbf{\% Gap to Ground State}}
\tabularnewline
\hline 
\textbf{Sites} & \textbf{Variables} & \textbf{Gilm} & \textbf{Elim} & \textbf{GRASP} & \textbf{Tabu} & \textbf{APT} \tabularnewline 
\hline 
\hline 
16 & 256 & 17.86 & 42.65 & 5.5 & 5.79 & 0 \tabularnewline
20 & 400 & 16.76 & 55.01 & 13.28 & 4.58 & 0 \tabularnewline
24 & 576 & 18.65 & 63.22 & 31.2 & 20.29 & 0 \tabularnewline
28 & 784 & 18.98 & 69.36 & 41.26 & 23.89 & 0 \tabularnewline
32 & 1024 & 19.76 & 77.4 & 48.51 & 33.32 & 0 \tabularnewline
36 & 1296 & 19.69 & 78.43 & 53.07 & 38.39 & 0 \tabularnewline
40 & 1600 & 18.83 & 78 & 55.27 & 46.52 & 0 \tabularnewline

\hline 
\end{tabular}
\caption{\label{table:runs} The gap to optimal configuration for various solvers on hard random QAP problems with known (planted) solutions \cite{drugan_generating_2015}. We obtained all the optimal solutions with our APT in less $10^{5}$ MC sweeps. Invoking nonlocal moves for these instances was not necessary due to high  efficiency of local APT. } 
\end{table}

In order to explore the performance separation of our nonlocal NMC against local APT for structured problems, we have benchmarked them on some intermediate-size industrial instances, with thousands of variables and millions of interacting terms, from QAPLIB \cite{burkard_qaplib_1997}, see App.~\ref{App:Industrial_QAP}. We observed that NMC achieves two orders of magnitude speedup over APT for obtaining relatively high quality solutions of certain QAP instances, as both were tuned by a ML-based hyperparameter optimization framework known as Vizier \cite{google_vizier}, see Fig.~\ref{fig:Vizier_sc32a_tho40}. Moreover, NMC demonstrated a robust performance such that on its worst worse runs it could beat the best APT runs over a wide range of time-scales, see Fig.~\ref{fig:sc32a_tho40}, which is remarkable considering its inherent nonequilibrium nature.

\section{Whitening procedure for low energy states}

\label{sec:whitening}

One important question regarding the power of NMC for sampling discrete configuration spaces is to quantify how many rare high quality solutions can be reached that are practically inaccessible with other solvers.  There are various ways to look at the distribution of solutions in configuration space, including generalized entropic measures, such as the Simpson diversity ~\cite{simpson_measurement_1949} and Renyi/Shannon entropies~\cite{spellerberg_tribute_2003}, or the Parisi order parameter \cite{MezardBook}. Recently, a new metric to quantify diversity of rare solutions in combinatorial optimization was introduced in Ref. \cite{mohseni_diversity_2021}. However, such measures do not directly quantify the size of frozen backbones in each solution.  Here, we employ an approach known as the \textit{whitening procedure}, originally proposed by Parisi \cite{parisi_local_2005}, that  finds a lower bound for the number of frozen variables for the ground state of $K$-SAT problems \cite{marino_backtracking_2016}, which we extend to low-energy states. For random 4-SAT near the computational phase transition, finding frozen cores of size $O(N)$ via the whitening procedure indicates the existence of rarely observed low-energy solutions residing beyond energy barriers that are widely believed to be exponentially hard to penetrate \cite{marino_backtracking_2016}. This is related to the concept of \lq\lq overlap gap property \rq\rq, or topological barrier in solution space of random structures, that has been recently used to explain algorithmic gaps, or absence of polynomial performance for a large class of algorithms in a regime between condensation phase transition and the actual computational phase transition \cite{gamarnik2021overlap}.    

The whitening procedure is a deterministic algorithm for a systematic inspection of all variables in a known solution. It assigns the \emph{white} label \lq\lq $\star$\rq\rq\  to any unfrozen variable, defined as those variables that can take different values without violating any clause in the Boolean formula. The procedure is iterative in nature and starts by inspecting one variable at a time and labeling it as a $\star$  only if all the clauses that the variable belongs to are either already satisfied by other variables or have another $\star$ variable. At each iteration, an increasing number of variables are labeled \emph{white} until we arrive at the steady state. At this fixed point any remaining variables must be frozen, since by construction such variables should belong to at least one clause that is only satisfied by this variable and contains no $\star$ variables. We note that the whitening procedure overestimates the number of \emph{white} or $\star$ variables, since it is essentially convexifing the cluster of solutions and thus provides a lower bound for the size of frozen backbones.

We generalize the whitening procedure to low-energy states by adding a single additional verification step:  any clauses containing a candidate frozen variable (those variables that have not been marked $\star$ at the steady state of the whitening procedure) must be satisfied by that variable. In other words, clauses that have been violated in a given low energy-state cannot report a frozen variable. This additional step increases the overwhitening nature of the procedure, as it could mark some extra $\star$ variables where otherwise would be considered frozen in the ground state. However, the advantage of our approach is that whenever we report a low energy state with a frozen variable, the result will be a conclusive outcome; i.e., we do not have false positives when reporting the existence of frozen variables.  

\begin{figure}[ht]
% \begin{subfigure}{\columnwidth}
 \centering
  \includegraphics[width=\columnwidth]{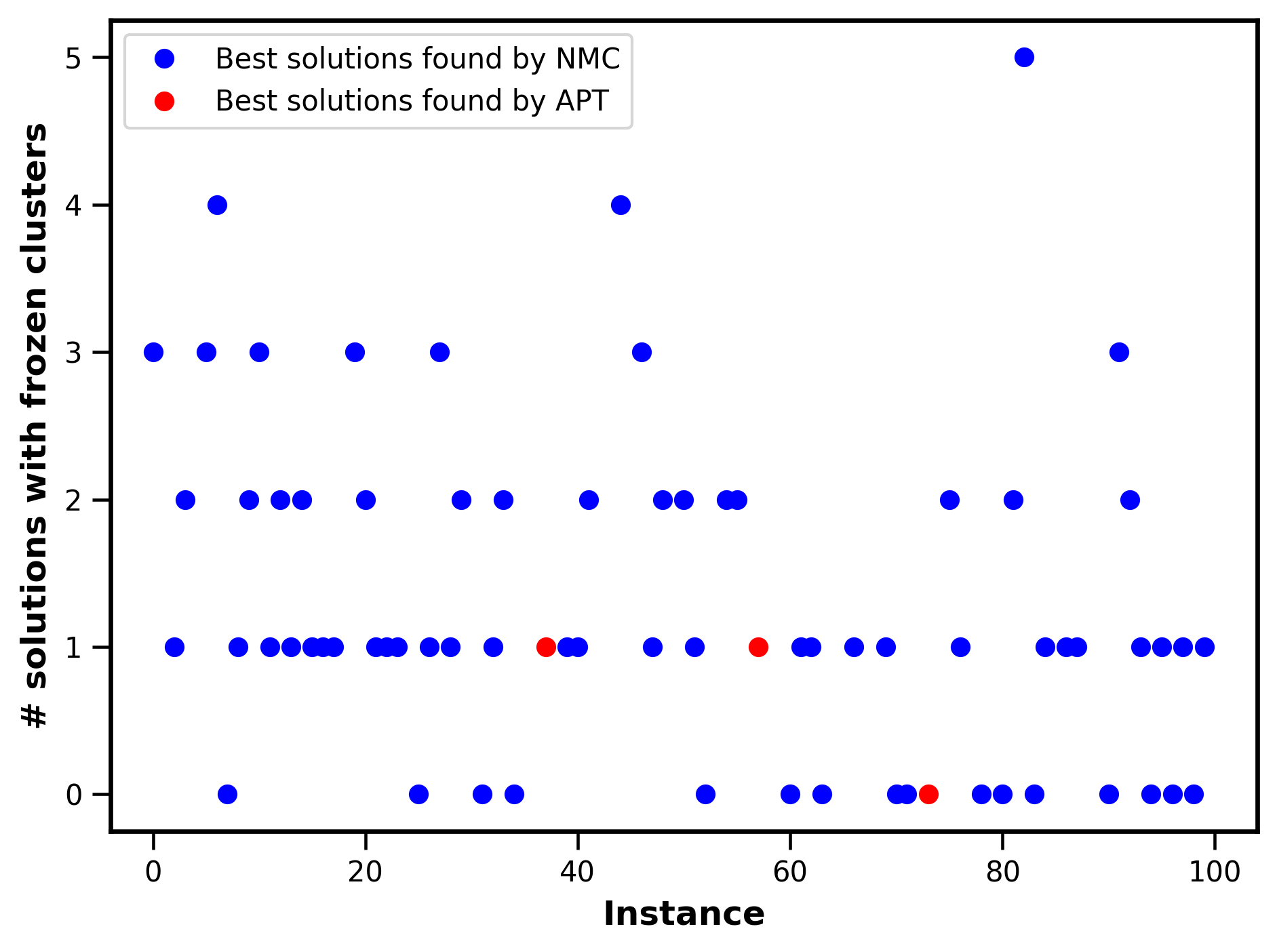}  
%   \caption{caption}
%   \label{fig:sub-first}
% \end{subfigure}
% \begin{subfigure}{\columnwidth}
%   \centering
  \includegraphics[width=\columnwidth]{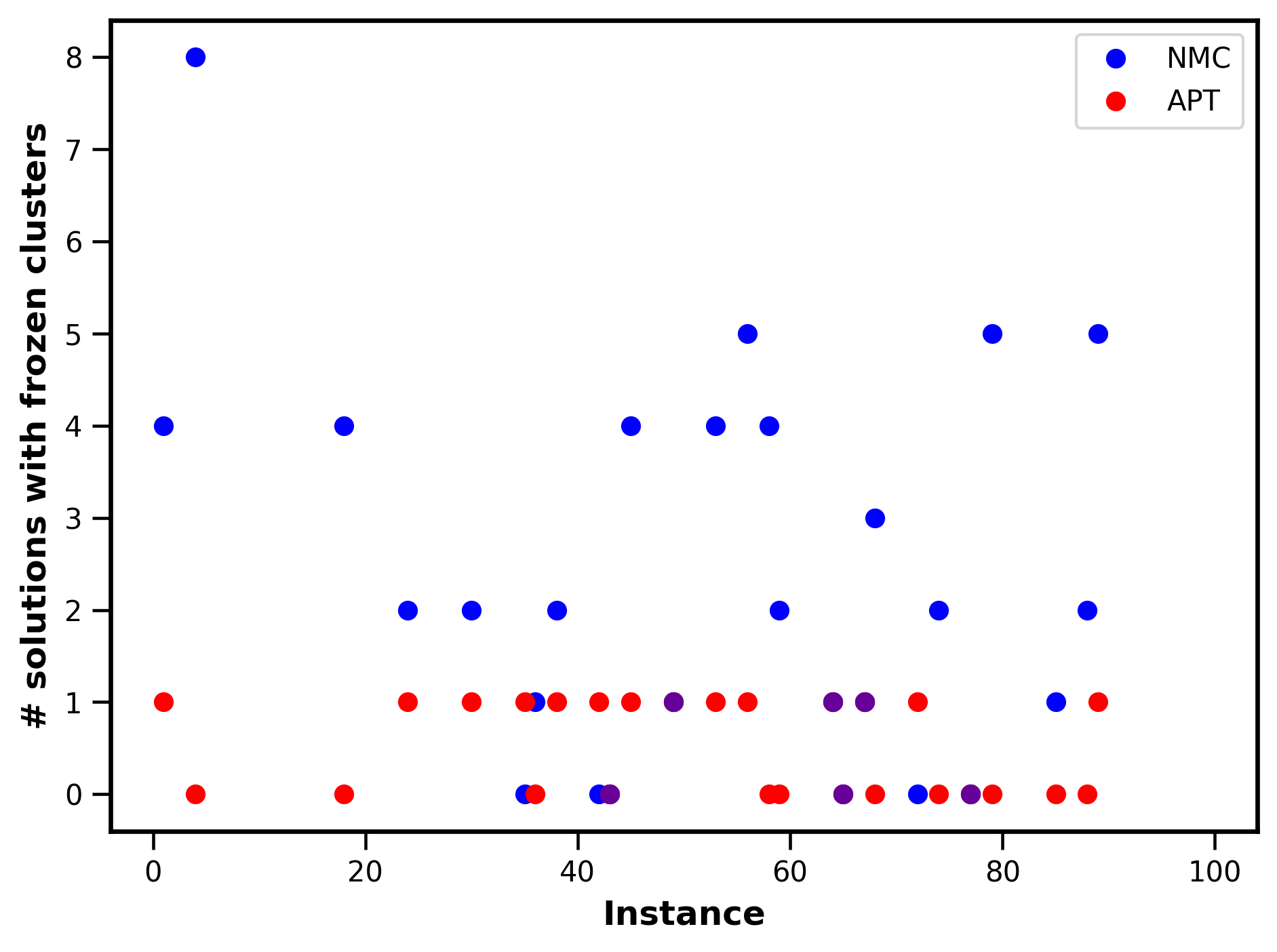}  
%   \caption{caption}
%   \label{fig:sub-second}
% \end{subfigure}
\caption{The number of best seen low-energy solutions within the approximation ratio of $2 \times 10^{-4}$ that contain frozen clusters of sizes $>4200$ variables found by either NMC and/or APT at about 50 repetitions. The instances are ordered according to their initial complexity. Upper panel: we found that for 54 instances the best low-energy solutions with frozen clusters  were only obtained by NMC (blue dots) in contrast to only 2 instances for APT (red dots). Lower panel: we note that even when NMC and APT get the same number of violations for 27 instances, there could be still considerable difference in the nature of solutions reported by each solver. For $60\%$ of such instances NMC could find multiple frozen solutions whereas APT did not find multiple frozen solutions even for a single instance.}
\label{fig:frozen}
\end{figure}

We used this whitening procedure to estimate the number of frozen variables in the low energy solutions found by our algorithm against those solutions found by the APT algorithm. In Fig.~\ref{fig:frozen} (top), we show 73 instances that either NMC or APT could find the best low-energy solutions, within the approximation ratio of $2 \times 10^{-4}$. We then computed the number of frozen clusters/backbones in such solutions that are involving at least $4200$ variables. We note that for 75\% of such instances the best seen solutions with frozen clusters were obtained by NMC versus about 3\% for APT. For 10 of these instances that were solved with very small number of violations (less than 4), we observed that the assignments with large number of frozen variables that were routinely found by NMC for 70\% of them, in just 4 repetitions, could not be found by APT even with O(1000) repetitions. 
In Fig.~\ref{fig:frozen} (bottom), we focus on the remaining 27 instances when the quality of best observed solutions were the same for both solvers. We note that APT did not report multiple solutions with frozen clusters for any single instance. In contrast, NMC could find multiple frozen solutions for 60\% of them, indicating a higher diversity, or better sampling, even when the performance of these two solvers matches with respect to the number of violations. We note that the lowest energy states found by a single run of SP (BSP) contained large frozen clusters for only one (three) instance(s) respectively.  

Thus, the NMC algorithm robustly reports many more solutions with a large fraction of frozen variables. This illustrates the existence of a new build-in mechanism for sampling the low energy manifold of the configuration space which is different in nature than other solvers studied in this work. This is a strong indication that NMC is effectively able to surpass certain barriers and enter into low-energy states that would be eventually inaccessible to other samplers.

\section{Conclusions and future directions}

\label{sec:conclusions}

In this work, we have demonstrated that the quantum-inspired Nonequilibrium Monte Carlo algorithm leads to effective shortcuts in configuration space, by unfreezing variables that are otherwise unresponsive to local moves at low temperatures. In this approach, we avoid the normal trade-off between exploration and exploitation via an adaptive interplay between two subroutines that are separately specialized for exploration and exploitation. More specifically, we use LBP on localized surrogate Hamiltonians to discover collective correlations among variables. We then build nonequilibrium inhomogenous MCMC for creating nonlocal updates to efficiently explore other possible basins of attraction in the configuration space. The interplay between these subroutines has the capacity for learning the correlations over discrete variables in different length scales.

We were able to get significant performance improvements over both generic and specialized solvers for QAP and random 4-SAT problems. In particular, for the 10\% of hardest random 4-SAT instances we observed one or two orders of magnitude improvements in the quality of solutions over  specialized solvers such as SP and BSP. The improvement in performance over local MC-based strategies, such APT, grows with the number of required MC sweeps leading to several orders of magnitude reduction in time-to-solution. This indicates that the larger and harder the problems are, the more benefit one will get from nonlocal moves. We quantified that the LBP subroutine only adds 5\% to 20\% overhead compared to local strategies and grows linearly with the size of the system.

There are several alternative algorithmic interpretations of our approach that might be worth discussing here. One can understand inhomogeneous MCMC as selectively flattening regions of the energy landscape that are related to the bottleneck energy barriers, without wiping out the other features in the rest of the energy landscape, as illustrated in Fig.~\ref{fig:NMC_animation}. In other words, inhomogeneous MCMC profiles with considerable maximum temperatures opens up saddle regions or crossways that can extend over a large area of configuration space and can be navigated with local moves. Alternatively, our approach can be considered as a new and generalized way of pruning the decision trees, in the same spirit of the CDCL-based SAT solvers \cite{biere_handbook_2021}, for seemingly unstructured optimization problems. One can envision our approach as a generalized probabilistic version of deterministic approaches that use the locality of the underlying structure for computational efficiency. For example, tensor-network contractions for 1D, 2D systems, or BP over tree-like structures lead to efficient factorizing of the joint probability distribution, or efficient ordering of summations over the relevant degrees of freedom,  by partitioning it over local regions. Essentially, we find a new approximate technique by inducing novel conditional independence of variables that are induced based on our instance-wise models of the backbone structures. This leads to a novel factorization over the localized subsystems, that are characterized by our surrogate backbones, even for high-dimensional strongly-disordered and highly-frustrated systems that do not have any apparent notion of locality. We note that such structures could be hidden to the known deterministic and probabilistic approaches, since the factorization of joint probability distributions is usually based on rather strong assumptions on the underlying symmetries, conditional independence and/or prior knowledge. 

Overall, we believe that our algorithm could have wide range of applications for combinatorial problems, mostly as a subroutine in conjunction with existing high performant solvers,  by essentially reducing the cardinality of the subset of worst-case instances, or reducing the algorithmic gap \cite{gamarnik2021overlap}. There are also significant challenges for learning and inference in structured graphical models with well-known computational bottlenecks related to the hardness of evaluation of marginal probability distributions or evaluation of partition functions. Thus, our approach could provide a new tool for approximate inference in Bayesian networks, Markov random fields, and training and inference in Boltzmann machines when known relaxation methods and variational techniques are ineffective.  Recently, there has been a considerable interest in deep learning models with a mixture of discrete and continuous variables \cite{oord2018neural}. We believe that our algorithm can be incorporated as a new computational primitive in such models for sampling over discrete data structures with underlying complex multimodal distributions.

\textbf{Acknowledgment.---} We would like to acknowledge useful discussions with Edward Farhi, Giorgio Parisi, John Platt, Vadim Smelyanskiy, and Jascha Sohl-dickstein. We would like to also thank David Applegate, Frederic Didier, Daniel Fisher, Pawel Lichocki, Jarrod McClean, Jon Orwant, Benjamin Villalonga, and Rif A. Saurous for feedback on this manuscript.

\bibliography{bib}

\appendix

\section{Adaptive Parallel Tempering}
\label{APT}

Models that have many local minima and high entropic barriers, such as spin glasses and hard optimization problems, are very hard to simulate. Conventional Monte Carlo methods with local updates get trapped within local minima and suffer from very long equilibration times. Cluster methods \cite{Swendsen_Wang87,Wolf89} do not work for such models because of frustration. Parallel tempering \cite{SwendsenWang1986,Geyer1991,HukushimaNemoto1996} is a generalization of the conventional Monte Carlo method. Many replicas are simulated at different temperatures. After a fixed number of conventional Monte Carlo sweeps, replica swaps are performed. This procedure is repeated many times. Parallel tempering has much faster convergence to equilibrium than conventional Monte Carlo. The acceptance probability to swap two replicas at adjacent temperatures $T_i$ and $T_j$ satisfies detailed balance and can be written as:
$$
  p_{ij} = \min \left\{1, \exp\left[ (\beta_i-\beta_j) (E(\beta_i)-E(\beta_j)) \right] \right\},
$$
where $\beta_i = 1/T_i$ is the inverse temperature and $E(\beta_i)$ is the configuration energy. 

The inverse temperature spacings $\beta_i-\beta_j$ should be chosen in such a way that the swap probabilities are not too high and not too low (typically, a value from the range $(0.2,0.3)$ is good enough). In general, a simple choice of geometrical schedule ($\beta_i=r\beta_{i-1}$ with fixed $r$) is not necessarily efficient in this respect. A temperature schedule that maintains the fixed swap probability that is independent of the temperature is much more efficient \cite{KoneKofke2005}. A number of methods have been developed to construct such a schedule adaptively \cite{Kofke2002,Rathore2005,Predescu2004,Predescu2005}. There are also methods that maintain denser temperature spacings in the vicinity of simulation bottlenecks such as phase transitions \cite{Katzgraber2006}. In this work, we are not exploring the algorithm space for adaptive versus non-adaptive PT. The main objective is to devise an adaptive strategy that allows for a simple hyper-parameter optimization and efficient benchmarking of nonlocal versus local PT. 

Here, we present a simple version of our adaptive algorithm. The basic intuition behind this algorithm is that the inverse temperature spacings should be small in the regions of high energy fluctuations, i.e. in the regions with large specific heat. Our algorithm automatically calculates the optimal number of replicas, temperature spacings between replicas, and minimum replica temperature. The adaptive temperature profile is calculated for each problem instance in the following prepprocessing procedure. Start with input parameters: maximum replica temperature $T_0$ (minimum inverse temperature $\beta_0$) and fixed parameter $\alpha$. Iteratively measure the energy variance $\sigma(\beta_i)^2$ (proportional to the specific heat) from
$$
  \sigma(\beta_i)^2=\left<E({\beta_i})^2\right>-\left<E({\beta_i})\right>^2
$$
by performing a number of Monte-Carlo sweeps at inverse temperature $\beta_i$ and calculate the next inverse temperatures $\beta_{i+1}$ from
$$
    \beta_{i+1} = \beta_i+\frac{\alpha}{\sigma(\beta_i)}.
$$
New $\beta$ values are generated until $\sigma(\beta_{\text{final}}) \leq \sigma_{min}$ for a specified $\sigma_{min}$ at which point the temperature is low enough for this problem instance and we do not need any more replicas at lower temperatures and the preprocessing step is finished. This algorithm maintains the replica swap probability $p$ that is more or less independent of temperature and loosely related to the parameter $\alpha$ via $p \approx e^{-\alpha^2}$.

Good values for $\alpha$ and $\sigma_{min}$ are easy to find. These values do not vary much between problem classes, for example, $\alpha=1.1$ is usually good enough for most problems and optimal values are rarely outside $[0.85, 1.25]$.

\section{Sampling localized replicas by Monte Carlo}

\label{App:localized_replica}

Here, we would like to demonstrate that in principle sampling over each localized surrogate Hamiltonian of our problem can be also done with standard Monte Carlo (MC) methods, but not as efficient or as reliable as LBP. This can be achieved as long as we stay within a single pure state where the ergodicity will not be an issue. In other words one samples by MC the measure proportional to
\begin{equation}
 \exp\left[-\beta H_\epsilon(\boldsymbol r)\right]   
\end{equation}
in order to estimate $\langle r_i \rangle$ and $\langle r_i r_j \rangle$, where $H_\epsilon(\boldsymbol r)$ is defined according to Eq. \ref{eq:explicitSurrogateHamiltonian}.

In this approach one can create $R$ clones, all initially identical to $\textbf{s}^{\star}$ and sample by MC the measure proportional to
\begin{equation}
  \exp\left[-\beta \sum_{a=1}^R H_\epsilon(\boldsymbol r^a)\right]
\label{eq:RclonesIndep}
\end{equation}
This increases statistics, as one can take empirical averages over all the clones. In the measure (\ref{eq:RclonesIndep}) the $R$ clones evolve independently, but all are constrained to have a given overlap with $\textbf{s}^{\star}$ (depending on $\epsilon$). Such a constraint induces in practice an effective coupling among the clones. Thus one can imagine to study a different measure where the clones are directly interacting
\begin{multline}
 P(\boldsymbol r^1,\ldots,\boldsymbol r^R) \propto \\ \exp\left[-\beta \sum_a H(\boldsymbol r^a) + \frac{\beta\gamma}{R-1} \sum_{a<b} \sum_i r_i^a r_i^b\right]
\label{eq:RclonesInter}
\end{multline}
Starting with all clones equal to $\textbf{s}^{\star}$ and $\gamma$ large enough, the MC should sample only the pure state $\textbf{s}^{\star}$ belongs to. When $\gamma$ is decreased, however, the clones can move away from $\textbf{s}^{\star}$ in a different form than the sampling measure (\ref{eq:RclonesIndep}) because in the latter clones are forced to stay at a given overlap $p$ with the reference configuration, while in measure (\ref{eq:RclonesInter}) the coupling $\gamma$ fixes the overlap between a pair of clones
\begin{equation}
 q_0 = \frac1N \sum_i \langle r^a_i \rangle \langle r^b_i \rangle \qquad \text{with} \quad a\neq b   
\end{equation}

Overall, we have 3 global parameters ($p,q_0,q_1$), given by the measure (\ref{eq:RclonesInter}), to describe the configurational space visited by the clones.
In principle, in the regime we are interested to sample the two measures (\ref{eq:RclonesIndep}) and (\ref{eq:RclonesInter}) should be equivalent.  This can be one additional criterion that the sampling method is faithful. 

In this work, we exclusively use LBP for discovering the correlations among the rigid variables as LBP guarantees linear scaling with respect to the surrogate Hamiltonians each forced to stay within a single pure state for sufficiently large $\epsilon$. Moreover, lack of convergence for LBP for too small values of $\epsilon$ provides evidence for leaving the basin of attraction. This feature will be absent if we evolve the clones by MCMC-based sampling.

\section{Loopy Belief propagation for k-local Hamiltonians over conjugate normal form}

\label{App:LBP_CNF}

Here, we provide the basic LBP equations for Surrogate Hamiltonians with k-local interactions over CNF:

\begin{align}
h_{i \to a} &= h_i + \epsilon s_i^\star + \sum_{b \in \partial i \setminus a} u_{b \to i}\\    
u_{a \to i} &= \beta^{-1} \text{arctanh}\left[
\tanh(\beta J_a) \prod_{j \in \partial a \setminus i} \tanh(\beta h_{j \to a})
\right],   \nonumber
\end{align}
where $\partial i$ is the set of factor nodes connected to vertex $i$ and $\partial a$ is the set of variable nodes connected to $a$. At convergence the LBP messages can be used to infer local marginals as follows
\begin{equation}
  \langle r_i \rangle = \tanh\bigg[\beta \Big(h_i + \epsilon s_i^\star + \sum_{a \in \partial i} u_{a\to i}\Big)\bigg]  \;.
\end{equation}
The high-order correlations are:
\begin{equation}
  \langle \prod_{i \in \partial a} r_i \rangle = \frac{\tanh(\beta J_a) + \prod_{i \in \partial a} \tanh(\beta h_{i\to a})}{1+\tanh(\beta J_a)\prod_{i \in \partial a} \tanh(\beta h_{i\to a})}  
\end{equation}

We set up the initial LBP messages for Ising as:
\begin{align}
h_{i\to a} &= \epsilon_i s_i^{\star} \\
u_{a \to i} &= \omega_a \prod_{j \in \partial a \setminus i} s^{\star}_j\;,
\end{align}
where $\omega_a$ is the weight for factor node $a$.

The LBP equations for the CNF formulation become:

\begin{equation}
h_{i\to a} = \epsilon_i s_i^{\star} + \sum_{b \in \partial i ^+ \setminus a} u_{b \to i}- \sum_{b \in \partial i ^- \setminus a} u_{b \to i},
\end{equation}
where $b \in \partial i ^+ \setminus a$ denotes the set of clauses in $\partial i$ agreeing with factor node $a$ on what values $i$ should take. Similarly, $b \in \partial i ^- \setminus a$ denotes the set of clauses in $\partial i$ disagreeing with factor node $a$ on what values $i$ should take. The messages from factor node to variables, $u_{a \to i}$, satisfy: 
\begin{equation}
u_{a \to i} = -\frac12 \ln\left[1 - (1-e^{-2\beta})\prod_{j \in \partial a \setminus i} \frac{1-\tanh h_{j\to a}}{2}\right] \nonumber\;.
\end{equation}
We set up the initial LBP messages for CNF as:
\begin{align}
h_{i\to a} &= \epsilon_i s_i^{\star} 
u_{a \to i} \\ &= -\frac12 \ln\left[1 - (1-e^{-2\beta})\prod_{j \in \partial a \setminus i} \frac{1-\tanh(\epsilon_i s_i^{\star})}{2}\right]\nonumber\;.
\end{align}

The marginal probability for a given variable i becomes:
\begin{multline}
\mu_i(r_i)=\frac12 \Large[1+ \\ r_i\tanh(\epsilon_i s_i^{\star} + \sum_{a \in \partial i ^+} u_{a \to i}- \sum_{a \in \partial i ^-} u_{a \to i})\Large],
\end{multline}
and local magnetization becomes:
\begin{equation}
\langle r_i \rangle =  \tanh\left(\epsilon_i s_i^{\star} + \sum_{a \in \partial i ^+} u_{a \to i}- \sum_{a \in \partial i ^-} u_{a \to i}\right)
\end{equation}
The joint probability distribution of variables in clause $a$ is
\begin{equation}
\mu_a(r_{\partial a})=\frac{1}{z_a}\omega_a(r_{\partial a}) \prod_{i \in \partial a} \frac{1-J^{a}_i r_i \tanh h_{i\to a}}{2},
\end{equation}
where $(J^{a}_1, J^{a}_2, ..., J^{a}_k)\in \{-1,+1\}^k$ are a set of constants that define the constraints in the k-local factor node $a$. Also, $\omega_a(r_{\partial a})$ is a weight for a factor node $a$ that equals 1 for all configurations $r_{\partial a}$ that satisfy this clause and $e^{-2\beta}$ for the single configuration that violates this clause. The normalization $z_a$ is defined as
\begin{equation}
\begin{split}
z_a = \sum_{r_{\partial a}} \omega_a(r_{\partial a}) \prod_{i \in \partial a} \frac{1-J^{a}_i r_i \tanh h_{i\to a}}{2} \\ = 1-(1-e^{-2\beta}) \prod_{i \in \partial a} \frac{1-\tanh h_{i\to a}}{2}
\end{split}\;.
\end{equation}
The high-order correlation function then becomes:
\begin{align}
%\begin{split}
\langle \prod_{i \in \partial a} r_i \rangle &= \sum_{r_{\partial a}\in \{-1,+1\}^k} \mu_a(r_{\partial a})\prod_{i \in \partial a}r_i \\ &= \frac{-(1-e^{-2\beta})\prod_{i \in \partial a} J_i^a \frac{1-\tanh h_{i\to a}}{2}} {1-(1-e^{-2\beta}) \prod_{i \in \partial a} \frac{1-\tanh h_{i\to a}}{2}}\;.
%\end{split}
\end{align}

\section{Generating disconnected clusters}

\label{App:alternate_backbones}

\subsection{Generating rigid clusters within local neighborhood of pure states}

In our first disconnected cluster grow strategy, we use the value of $p = \frac1N \sum_i s^{\star}_i \langle r_i \rangle$  as a guiding principle to set the size of the clusters of correlated spins. By construction, $p$ quantifies the average distance of surrogate Hamiltonians variables that are sampled via LBP from the reference configuration. In general, certain (typically small) threshold value of p could exist in which beyond that the correlations cannot be reliably estimated by LBP. This imposes a natural upperbound on the number of variables that can be possibly included into rigid clusters. We define $p^*$, which is the target value of $p$ using $p^* = 1 - 2L^*/N$, where $L^*$ is a target cluster size; e.g., $N/4$. We start with sufficiently large value of $\lambda$ and decrement it as we perform LBP iteratively until $p$ becomes sufficiently close to $p^*$ or until $p$ reaches a constant fixed point with respect to $\lambda$. We then grow the cluster by adding the top nearest-neighbor correlated pairs of spins until it reaches the size set by $p$.

\subsection{Generating disconnected rigid clusters while LBP converges}

In this strategy, we first find the smallest possible global $\lambda$ for LBP iterations that still converge within some desired convergence precision, then we calculate the marginals and two-point correlations. We grow the clusters by including all pairs of spins with effective couplings above a specified correlation threshold. Here, we start off with a sufficiently large value of the $\lambda$  and gradually reduce its rescaling factor. In each step, we perform LBP iteratively until the LBP messages do not converge or $\lambda$ reaches a preassigned lower bound. We then pick the smallest $\lambda$ for which the LBP messages had already converged. We then calculate the nearest neighbor correlations and nearest neighbor effective couplings for all spins and re-scale the effective couplings by $J_{ij}$. We create the cluster by adding all spin pairs with re-scaled effective coupling above the correlation threshold.

\section{Testing NMC primitives}
\label{App:other_problems}

In this section, we provide a few examples on versatility of NMC algorithm as a generic solver. We observed that all of our algorithmic primitives for discovering frozen variables, growing meaningful backbones, and the construction of nonequilibrium inhomogeneous MCMC work reliably on problems with different dimensionality and distributions of interactions.

%%%%%%%%%%%%%%%%%%%%%%%%%%%%%%%%%%%%%%%%%%%%%%%%%%%%%%%%%%%%%%%%%%%%%%%%%
\begin{figure}[t]
    \centering
    \includegraphics[width=\columnwidth]{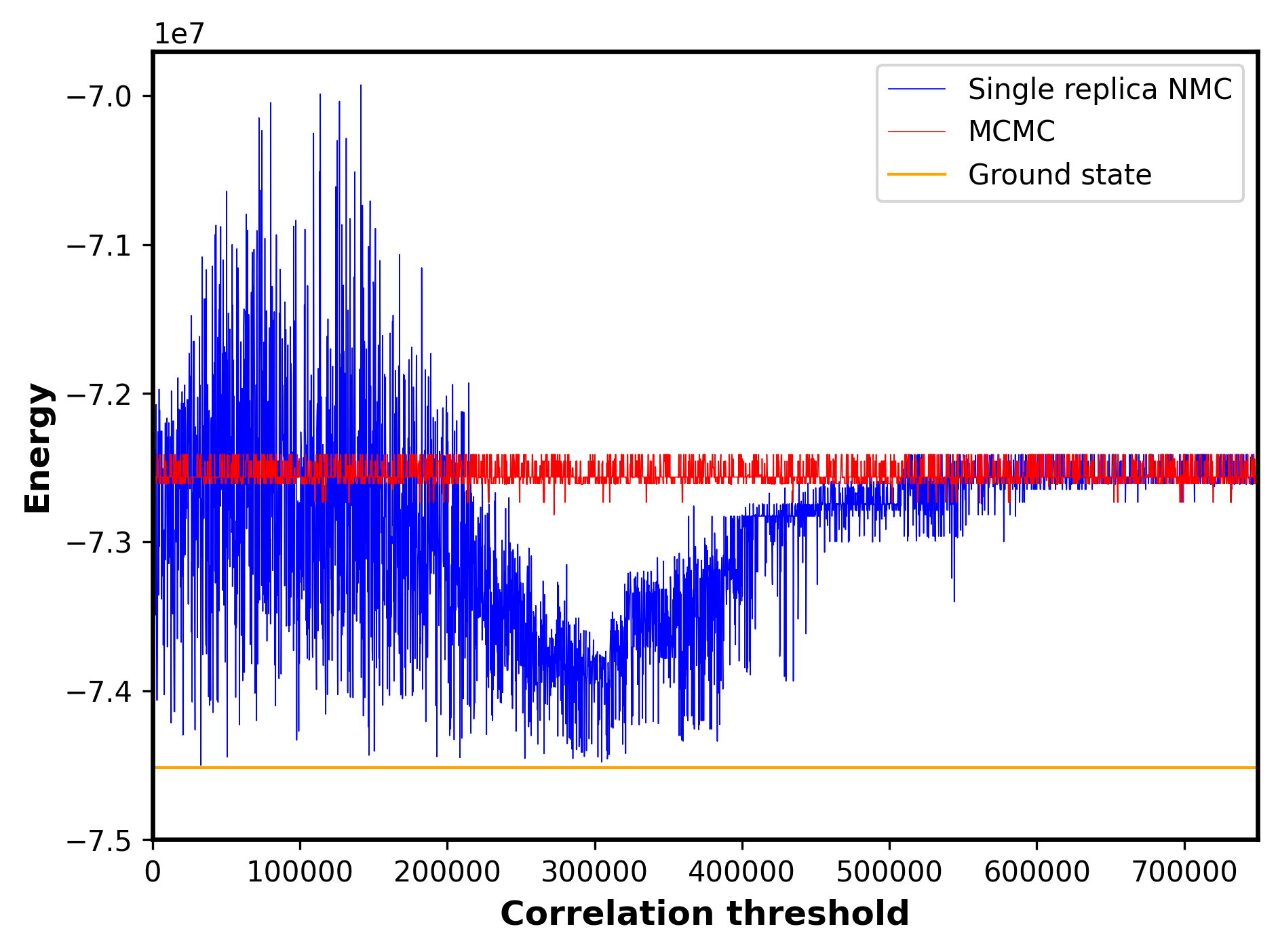}
    \caption{Best solutions obtained for a Max-Cut instance with 380 variables for a single replica standard MCMC (red curve) with 200k sweeps over 4000 repetitions. This can be compared with best energy obtained after using a seed from standard MCMC after 100k sweeps and then performing 100k inhomogeneous MCMC induced by surrogate backbones calculated as a function correlation threshold (blue curve). We see a wide range of optimal correlation threshold. When the correlations threshold are too high (surrogate backbones are too small) we recover standard local MCMC as expected.}
    \label{fig:Max_Cut}
\end{figure}
%%%%%%%%%%%%%%%%%%%%%%%%%%%%%%%%%%%%%%%%%%%%%%%%5%%%%%%%%%%%%%%%%%%%%%%%

%%%%%%%%%%%%%%%%%%%%%%%%%%%%%%%%%%%%%%%%%%%%%%%%%%%%%%%%%%%%%%%%%%%%%%%%%
\begin{figure}[t]
    \centering
    \includegraphics[width=\columnwidth]{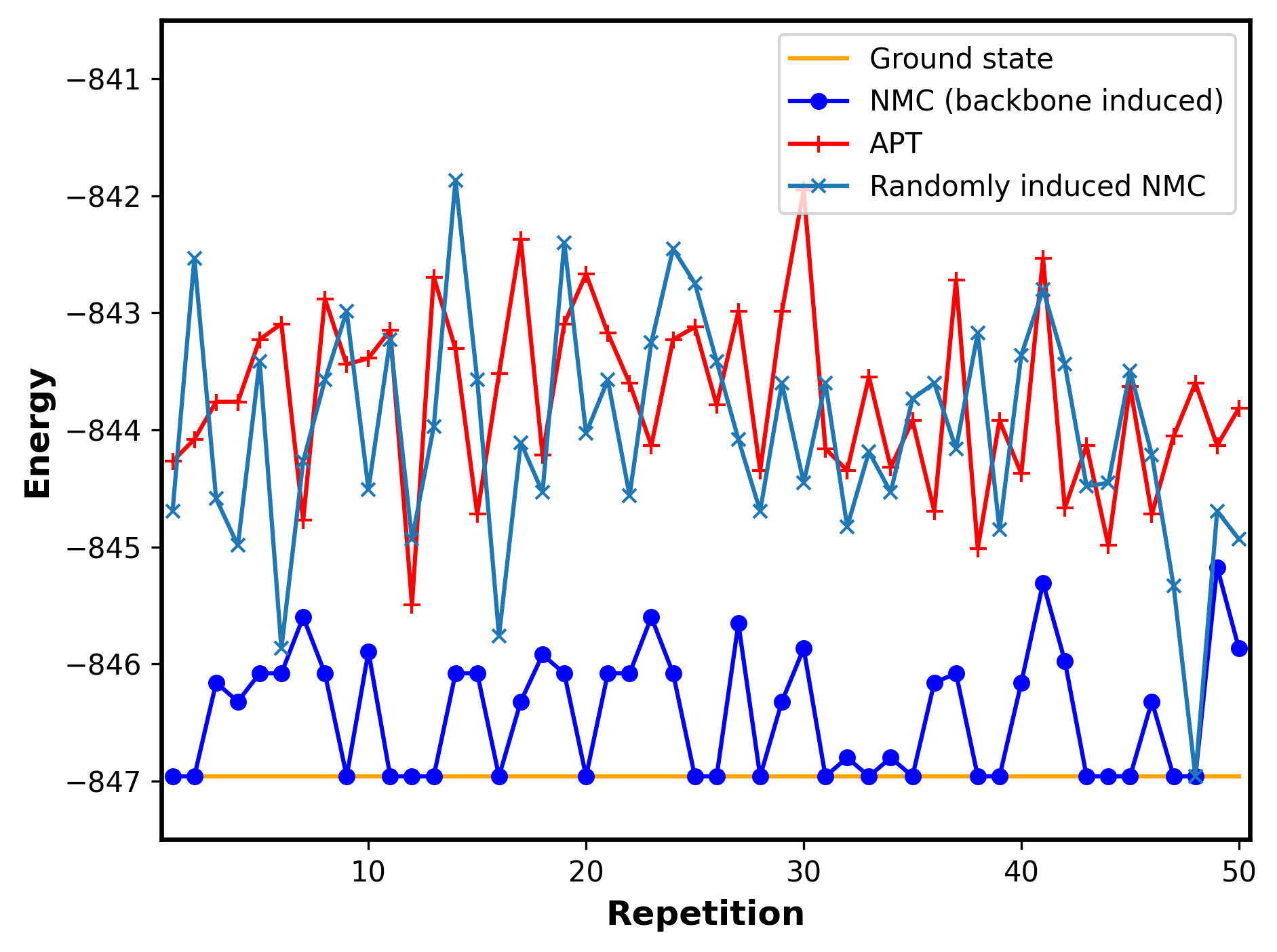}
    \caption{Best residual energy found with various MC-based solvers for a structured Chimera instance from Ref~\cite{rams_tensors_2021} with a total of $10^{7}$ sweeps over 50 repetitions. We observe no improvement over standard APT when using an inhomogeneous APT schedule that are induced by some random clusters. In contrast, if we employ an inhomogeneity that are devised based on surrogate Hamiltonian backbones then much lower energy manifold, which contains the ground state, can be sampled routinely.}
    \label{fig:Chimera_full_APT}
\end{figure}
%%%%%%%%%%%%%%%%%%%%%%%%%%%%%%%%%%%%%%%%%%%%%%%%5%%%%%%%%%%%%%%%%%%%%%%%%%

We studied the performance of our subroutines on some random and structured instances generated over the Chimera graph, with a quasi-2D geometry. We also tested these subroutines on some of the random and structured Max-Cut instances. For different problem classes, we observed that the size of surrogate Hamiltonian backbones does not suddenly percolate as a function of the correlations and shows a power law distribution. Thus, one can use correlation threshold as a new hyperparameter to explore emerging correlations in various length scales by tuning this as an effective control knob.

In Fig.~\ref{fig:Max_Cut}, we investigate the robustness of the correlation threshold cutoff for a structured Max-Cut instance with about 380 variables over a single MCMC replica. The best residual energy obtained by inhomogeneous MCMC can be substantially improved with optimal values of correlation threshold as a free hyperparameter. This demonstrates that our approach is not ultra sensitive to correlation threshold cutoff since a significant performance separation between NMC versus the standard MCMC can be observed over a wide range of correlation threshold. At the two extreme limits of very large or very small correlation threshold cutoff one can recover the performance of standard (local) MCMC and random cluster update strategies.  

Fig.~\ref{fig:Chimera_full_APT}, show the results of our study on one of the structured Chimera instances from Ref. \cite{rams_tensors_2021} for a total $10^{7}$ sweeps. It can be seen that, except one run, there is not significant performance improvement for various repetitions of standard APT versus an alternative inhomogeneous APT, in which the inhomogeneity schedule is simply induced by random clusters. However, the performance of NMC, with inhomogeneoity driven by the surrogate backbones, generate significantly lower energy states (founding ground state in many runs), such that in its worse runs outperforms best runs of both standard APT and randomly-induced inhomogeneous APT.

\section{NMC with annealed correlation thresholds}

\label{App:adiabaticNMC}

For solving an ensemble of hard random 4-SAT problems at the computational phase transition, we develop a quantum-inspired approach and quasi-adiabatically anneal the values of the correlation threshold cutoffs from low values (where the maximum size of emergent backbones would be around $N/2$) to high values near unity (where the maximum size of clusters would be in the single digits). See Algorithm \ref{algorithm2}. The range of correlation thresholds values for this adiabatic transitions can be obtained with a comparison of the estimated correlations with the largest energy-scale of system or with a few preprcessing trials. In this quasi-adiabatic strategy, we invoke aggressive nonlocal moves with very large fluctuations in the configuration space at the beginning phase of the algorithm, in analogy with large Hamming distance variations in quantum annealing due to initially large quantum fluctuations in the annealing schedule. These nonlocal moves can be implemented either by increasing the temperature within the backbone, as we will explain in Sec. \ref{sec:balance} and also shown in the Fig. \ref{fig:NMC_animation} (d) and (f), or with a collective spin-flip update over the backbone (shown in the Fig. \ref{fig:NMC_animation} (c)). We note that similar aggressive collective-like updates can also be realized via an infinite temperature inhomogeneity similar to those finite temperatures inhomogeneity that are depicted in Fig. \ref{fig:NMC_animation} (d) and (f). 

We then adiabatically became more conservative, by increasing the value of the correlation threshold monotonically which consequently reduced the size of clusters. We spend a considerable amount of time near the critical point where the size of clusters exhibit a power-law distribution. Eventually the algorithm phase out all nonlocal moves and ends up with pure local MCMC; e.g., for the last $10\%$ to $20\%$ of the overall run time, focusing on exploitation instead of more exploration resulting in global steady state without strictly satisfying a detailed balance condition.

%\newpage

\begin{minipage}{1\linewidth}
\begin{algorithm}[H]
\SetAlgoLined
\SetInd{1em}{1em}

\While{good solutions not found}{

\textbf{Replica exchange MC}: Adaptive homogeneous replica-exchange MC on the entire problem.

\For{replicas at low temperatures}{

\textbf{Generate seeds}: Find a low energy state as a seed solution, $s^{*}$.

\For{each seed}{

\textbf{Build localized problems}: Construct a localized surrogate Hamiltonian around the neighborhood of a seed solution.

\textbf{Infer correlations}: Use efficient approximate inference techniques, such as LBP, to estimate marginals over the localized surrogate problem.

}

\For{correlation thresholds in annealing range}{

 \textbf{Grow backbones}: construct surrogate backbones in different length scales.

\While{arriving at a steady state}{

\textbf{Nonlocal exploration:} Inhomogeneous Monte Carlo on backbone subproblem by conditioning over non-backbone variables.

\textbf{Local exploitation:} 
Inhomogeneous Monte Carlo on non-backbone subproblem by conditioning over backbone variables.

\textbf{Unlearning phase:} Perform homogeneous MCMC sampling on full problem to repair topological defects at the backbone boundaries.
}
}
}
} 
 \caption{ \textsc{Nonequilibrium Monte Carlo (NMC) with annealing subroutines}}
 
 \label{algorithm2}
\end{algorithm}
\end{minipage}

%\newpage

\section{The Survey Propagation Algorithm and the Backtracking Survey Propagation}
\label{App:SP-BSP}

The Survey Propagation algorithm (SP) is a heuristic message passing algorithm, developed by Mezard, Parisi, and Zecchina \cite{mezard_analytic_2002} from the assumption of one-step replica symmetry breaking and the cavity method of spin glasses. SP works on the factor graph with underlying CNF formula. For larger $N$, SP is conjectured to work better as it runs over locally-tree like factor graphs, and cycles into the graph are $O(\log N)$. A detailed description of the Survey Propagation algorithm can be found in \cite{mezard_analytic_2002, braunstein2005survey}, but here we summarize the main results. 

Broadly speaking, SP exchanges messages between the $N$ variables and $M$ clauses in order to guess the value that each variable needs to be set for satisfying all clauses. More precisely, a message of SP, called a survey, passed from one function node $a$ to a variable node $i$ (connected by an edge) is a real number $\eta_{a\to i} \in [0, 1]$. The messages have a probabilistic interpretation under the assumption that SP runs over a tree-like factor graph. In particular, the message $\eta_{a\to i}$ corresponds to the probability that the clause $a$ sends a warning to variable $i$, telling which value the variable $i$ should adopt to satisfy itself \cite{parisi2003probabilistic, maneva_new_2007}.

The iterative equations of SP are:
\begin{equation}
\label{algoSP}
\begin{split}
&s_{j\to a}^{\mp}=\left[1- \prod_{b \in \partial_{ja}^{\mp}} (1-\eta_{b\to j}) \right]\prod_{b \in \partial_{ja}^{\pm}} (1-\eta_{b\to j})\\
&s_{j\to a}^{0}=\left[\prod_{b \in \partial_{j} \setminus a} (1-\eta_{b\to j}) \right]\\
&\eta_{a\to i}=\prod_{j \in \partial_{a} \setminus i} \left[ \frac{s_{j\to a}^{-}}{s_{j\to a}^{-}+s_{j\to a}^{+}+s_{j\to a}^{0}}\right];
\end{split}
\end{equation}
where the symbol $\partial_a$ defines the set of variables nodes connected with the functional node $a$, i.e., the variable in clause $a$, and the symbol $\partial_i$ defines the set of functional nodes connected with the variable node $i$, i.e., the set of clauses where the literal $x_i$, or $\overline{x}_i$, appears. The cardinality of the set $\partial_i$ is the degree of a variable node $i$, i.e., the number of links connected to it, and is defined with $n_i$. The set $\partial_i$ is composed of two  subsets, namely $\partial_i^+$ that contains the functional nodes where the variable node $i$ appears not negated, and $\partial_i^-$ that contains the functional nodes where the variable node $i$ appears negated. Obviously, the relation $\partial_i= \partial_i^+ \cup \partial_i^-$ holds.

With the symbol $\partial_{ia}^{+}$ (respectively $\partial_{ia}^{-}$) we define the set of functional nodes containing the variable node $i$, excluding the functional node $a$ itself, satisfied (respectively not satisfied) when the variable $i$ is assigned to satisfy clause $a$. In other words, if the literal $x_i$ is not negated in the clause $a$, then the $\partial_{ia}^{+}$ is the set of functional nodes containing the variable node $i$, excluding the functional node $a$ itself, where the literal $x_i$ appears not negated, while $\partial_{ia}^{-}$ is the set of functional nodes containing the variable node $i$, where the literal $\overline{x}_i$ appears negated. In contrast, if the variable $i$ is negated in the clause $a$, then the $\partial_{ia}^{+}$ is the set of functional nodes containing the variable node $i$, excluding the functional node $a$ itself, where the literal $\overline{x}_i$ appears negated, while $\partial_{ia}^{-}$ is the set of functional nodes containing the variable node $i$, where the literal $x_i$ appears not negated.
 
SP is a local algorithm that extracts information on the underlying graph of a CNF formula. As input, it takes a CNF formula of a random SAT Problem, and it performs a message-passing procedure to obtain convergence of the messages. More precisely, we are given a random initialization to all messages, and at each iteration, each message is updated following eq. \eqref{algoSP}. SP runs until all messages would satisfy a convergence criterion. This convergence criterion is defined as a small number $\epsilon$ such that the iteration is halted at the first time $t^*$ when no message has changed by more than $\epsilon$ over the last iteration. If this convergence criterion is not satisfied after $t_{max}$ iterations, SP stops and returns a failure output. Once a convergence of all messages $\eta_{a \to i}$ is found, SP computes the marginals for each variable $i$:
\begin{equation}
\label{SID}
S_i^-=\frac{\pi_i^-(1-\pi_i^+)}{1-\pi_i^+\pi_i^-},\\
S_i^+=\frac{\pi_i^+(1-\pi_i^-)}{1-\pi_i^+\pi_i^-},\\
S_i^0=1-S_i^--S_i^+,
\end{equation}  
where:
\begin{equation}
\pi_i^{\pm}=1-\prod_{b\in \partial_i^{\pm}}(1-\eta_{b \to i}).
\end{equation}
The SP marginal $S_i^{+}$ ($S_i^{-}$) represents the probability that the variable $i$ must be forced to take the value $x_i=1$($x_i=0$), conditional on the fact that it does not receive a contradictory message, while $S_i^{0}$ provides the information that the variable $i$ is not forced to take a particular value. 

Once all the SP marginals have been computed, the decimation strategy can be applied. This algorithm is called Survey Inspired Decimation (SID). Decimating a variable node $i$ means fixing the variable to $1$ or $0$ depending on the SP marginals, removing all satisfied functional nodes and the variable node $i$ from the factor graph, and removing all the literals into the clauses that have not been satisfied by the fixing. How to choose the variable node $i$ to decimate? The answer is simple, just selecting a variable with the maximum bias $b_i=1-\min(S_i^-, S_i^+)$. Decimated the variable node $i$, the SP algorithm tries to find out a new state of convergence and uses decimation again until one of these three different outcomes appears: (i) a contradiction is found, then the algorithm returns exit failure; (ii) SP does not find a convergence, then the algorithm returns exit failure; (iii) all the messages converge to a trivial fixed point, i.e., all the messages are equal to $0$. In all these cases, the algorithm calls WalkSAT, either because the residual formula should be easy to treat (case iii) or because the lack of convergence or the contradiction can be due to the formula acquiring non-random structure upon decimation, and keep running SP would not lead to any improvement (cases i and ii). WalkSAT then tries to solve the residual formula and eventually builds the complete solution of the problem. For the numerical analysis displayed above, we modified step (ii) of the algorithm as: (ii) SP does not find a convergence after $t_{max}=1024$ iterations, then calls WalkSAT, which builds a low energy assignment of the problem. 

This algorithm has low complexity. Each SP iteration requires $O(N)$ operations, which yields $O(N t_{max})$, where $t_{max}$ is the maximum time allowed for finding a convergence, i.e., a big constant. In the implementation described above, the SID has a computational complexity of $O(t_{max} N^2 \log N )$, where the $N\log N$ comes from the sorting of the biases. This can be reduced to $O(Nt_{max}(\log N)^2)$ by noticing that fixing a single variable does not affect the SP messages significantly. Consequently, SP can be called every $N\delta$ decimation step by selecting a fraction of variables at each decimation step. The efficiency of the algorithm recalled above can be improved by introducing a backtracking strategy. We refer to \cite{marino_backtracking_2016} and references therein for a complete explanation of this strategy. Here, we summarize it.

The fact the SID algorithm assigns each variable only once is clearly a strong limitation, especially in a situation where correlations between variables becomes strong and long-ranged. In difficult problems it can easily happen that one realizes that a variable is taking the wrong value only after having assigned some of its neighbours variables. The backtracking survey propagation (BSP) algorithms \cite{parisi2003backtracking} tries to solve this kind of problematic situations by introducing a new backtracking step, where a variable already assigned can be released and eventually re-assigned in a future decimation step.  It is not difficult to understand when it is worth releasing a variable. The strategy is to release the variables with a value of the bias $b_i$ which is smaller than those variables already fixed. 

The BSP algorithm then proceeds similarly to the SID algorithm above described. It applies either a step of decimation or a step of backtracking on a fraction of variables, after the iterative solution of the SP equations are obtained.  The choice between a decimation or a backtracking step is taken according to a stochastic rule with a parameter $r\in [0,1)$. This parameter represents the ratio between backtracking steps to decimation steps. Obviously for $r=0$, we recover the SID since no backtracking step is ever done. Increasing $r$ the algorithm becomes slower by a factor $1/(1- r)$, because variables are reassigned on average $1/(1 - r)$ times each before the BSP algorithm reaches the end, but its complexity remains at most $O(N (\log N)^2)$ in the problem size \cite{marino_backtracking_2016}.

The BSP algorithm introduces only a new decimation strategy, and therefore it can stop for the same reasons the SID algorithm does: either the SP equations cannot be solved iteratively or the generated subproblem has a contradiction. In the case all the messages converge to a trivial fixed point, the subproblem can be given to WalkSat which solves the residual formula and builds the complete solution of the problem. Again, for our numerical analysis the modification on step (ii) of SID holds.

\section{Instance-wise hardness: Large/Small performance fluctuations for BSP/NMC algorithms}
\label{App:BSPperf}

In this section, we quantify the instance-wise performance of BSP based on the entropy of clusters of solutions for the original formulas and the size of residuals formula when it fails to converge. We then compare and contrast the BSP vast dispersion in the quality of solutions with highly robust and reliable performance of NMC across various repetitions over a wide range of instance-wise complexity.

As explained in App.~\ref{App:SP-BSP} the SID algorithm is essentially a deterministic algorithm as its output depends mildly on the few parameters required, e.g.\ the fraction of variables decimated at each step or the maximum number of SP iterations. So, running SID once would be enough to make a fair comparison with our new NMC algorithm. On the contrary, the BSP algorithm follows a more stochastic rule. Indeed at each step the decision between decimating a fraction of most biased variables or backtracking, removing the assignment to the least bias variables, is taken according to a random number.
This means each run of the BSP algorithm can follow a different path in the assignment of variables searching for a solution and this in turn can lead to very different outputs. Here, we analyze the performance of the BSP algorithm, focusing in particular to the variations between different instances and different runs in the same instance.

\begin{figure}
    \centering
    \includegraphics[width=\columnwidth]{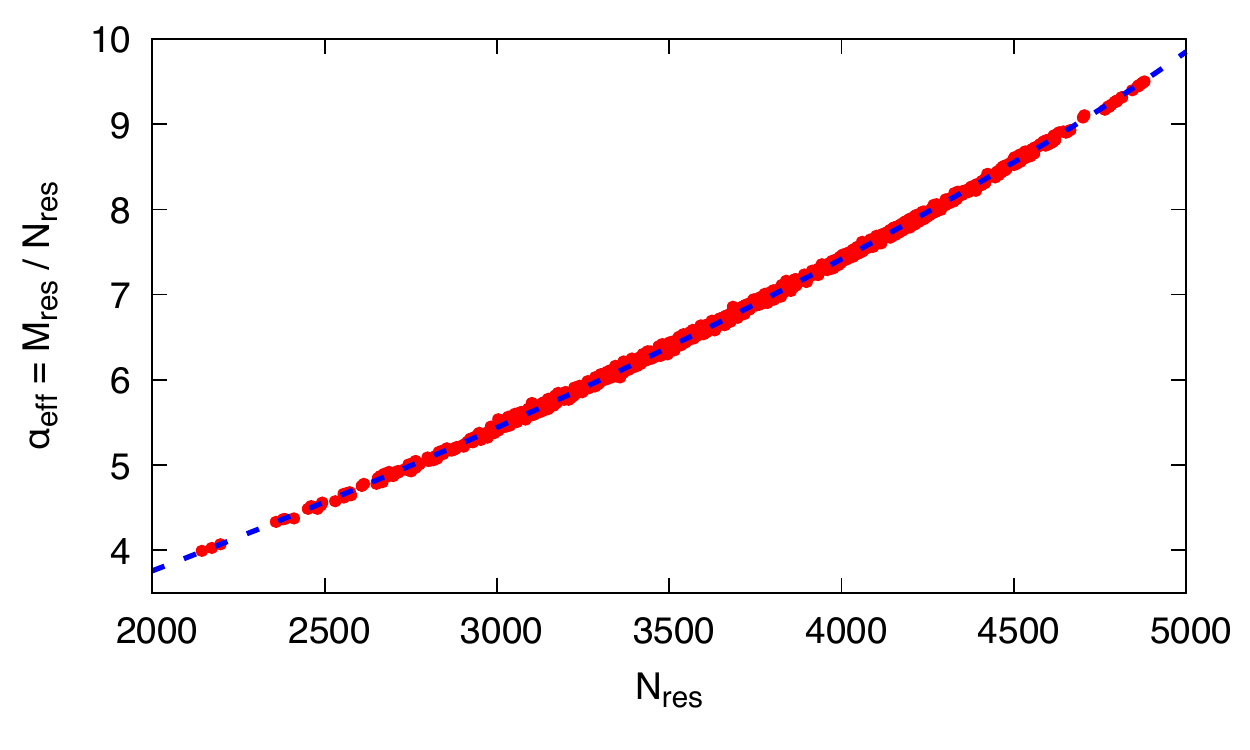}
    \caption{The residual formula after decimation by BSP has $\Nres$ variables and $\Mres$ clauses. These two quantities are strongly correlated as shown in the figure, where data for all the 100 instances and all the runs have been used together. The dashed blue line is just a guide for the eyes. The hardness of the residual formula strongly depends on its size $\Nres$ and this affects the success probability of the WalkSAT solver which is called with the residual formula as input.}
    \label{fig:res}
\end{figure}

In Ref.~\cite{marino_backtracking_2016} the BSP algorithm was used to solve extremely large formulas, that were essentially at the thermodynamic limit, and no relevant fluctuations between instances and/or runs were observed. 
On the contrary, in the present work we are focusing on formulas which their sizes are comparable to typical large real-word applications, yet they are not strictly at the thermodynamic limit. Moreover, these instances are generated deep into the rigidity phase (i.e.\ the range of $\alpha$ values where frozen variables dominate typical solutions in the large $N$ limit). For these reasons we expect much larger fluctuations in the behavior of the BSP algorithm which have not yet been previously explored. 

The most relevant parameter in the BSP algorithm is the ratio between backtracking and decimation moves that we fix to $r=0.999$ in order to make contact with previous studies. As in the SID algorithm the other parameters are not so crucial: we fix the fraction of variables to decimate/backtrack at each step to $0.125\%$ and the maximum number of SP iterations to $1024$.

\begin{figure}
    \centering
    \includegraphics[width=\columnwidth]{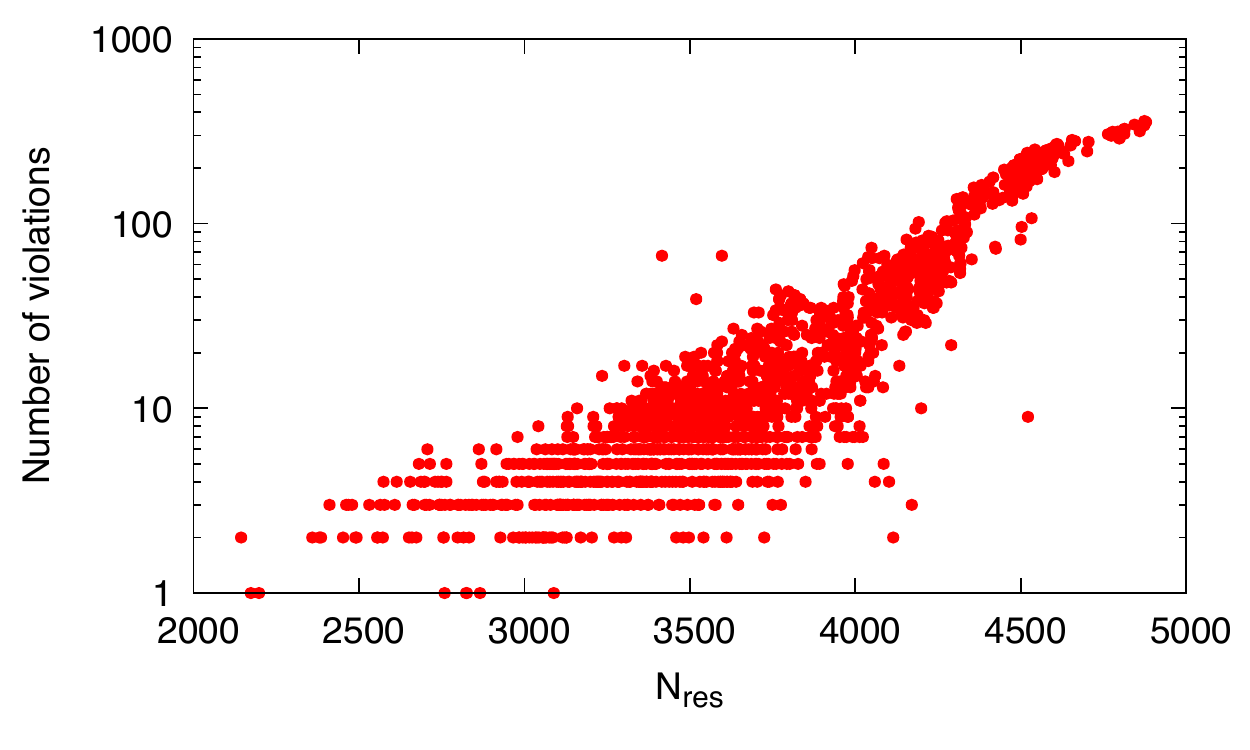}
    \caption{The minimal number of violations found running WalkSAT on the residual formula as a function of the number of variables $\Nres$ in the residual formula. It is clear that the success of WalkSAT in finding a low energy configurations strongly depends on the ability of BSP of returning a residual formula small enough.}
    \label{fig:ener}
\end{figure}

In the BSP algorithm, after a certain number of decimation/backtracking rounds, eventually SP stops converging and the algorithm generates a \emph{residual formula} to be passed as input to the WalkSAT solver in the search for the optimal configuration or even a solution. The residual formula has $\Nres$ variables and $\Mres$ clauses. The two quantities are strongly dependent as shown in Fig.~\ref{fig:res} where we plot  the data from all the BSP runs on the 100 different instances with $N=5000$ and $\alpha=9.884$. The effective ratio $\alpha_\text{eff}=\Mres/\Nres$ becomes smaller for smaller $\Nres$, indicating that formulas with a smaller $\Nres$ are expected to be easier to solve. This is clear from the data shown in Fig.~\ref{fig:ener} where we are plotting the smallest number of violations or minimal energy found by running WalkSAT on the residual formula returned by BSP. In practice low energy configurations can be reached only if BSP outputs a small enough residual formula. However, BSP can fail spectacularly when there is a large residual formula. We observe BSP can report a significant number of violations or final energy, up to two orders of magnitude larger than the typical lowest energies.

Here, we would like to address two key questions: how often does BSP return a residual formula that is too large such that the final solutions are very low quality?  Is this poor performance of BSP correlated with the intrinsic hardness of the instance? Even if all instances have been generated with $N=5000$ and $\alpha=9.884$, we expect measurable variations in the actual hardness in solving these instances: indeed being very close to the SAT/UNSAT threshold and deep into the rigidity phase, even a small fluctuation in the structure of the instance may lead to visible changes in hardness.

\begin{figure}
    \centering
    \includegraphics[width=\columnwidth]{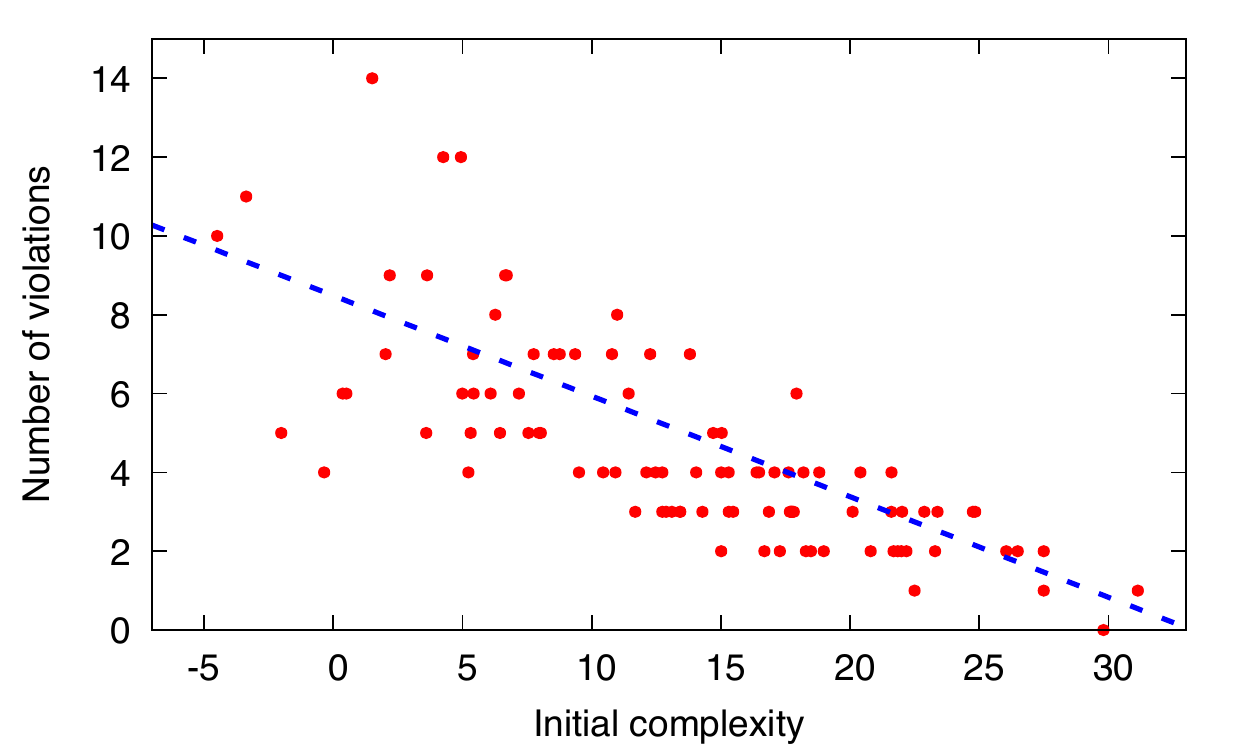}
    \caption{The complexity computed by SP on the original formula (before decimating it) correlates very well with the best final energy reached by running the BSP algorithm. So we can use the initial complexity as a proxy for the hardness of a given instance.}
    \label{fig:complex}
\end{figure}

In general, there is no explicit ``measure of hardness'', but here we can resort to an empirical one. For each instance, we consider the BSP run that returned the best final energy. We then plot this energy as a function of the complexity of the original formula computed from the first fixed point reached by SP. 
We report the result in Fig.~\ref{fig:complex} which exhibits a strong correlation between this two physical quantities. The advantage of this notation of complexity is that it can be computed on the original formula, before any decimation taking place. This notation of complexity, $\Sigma$, is related to the number of clusters of solutions $\mathcal{N}_{\text{clu}}$ according to Eqs:  \ref{Eq:complexity1}, \ref{Eq:complexity2}, and \ref{Eq:complexity3}.
\begin{comment}
\begin{equation}
\begin{split}
   &\Sigma = \log(\mathcal{N}_{\text{clu}})= \sum_{i=1}^{N} {\Sigma_i} + \sum_{a=1}^{M}(1-|\partial_a|){\Sigma_a};\\ 
   \end{split}
\end{equation}
where
\begin{equation}
\begin{split}
   &\Sigma_a=\log(1-\prod_{j \in \partial_{a}} \eta_{j \to a});\,\,\, \Sigma_i=\log(1-\pi^{+}_{i}\pi^{-}_{i})
\end{split}
\end{equation}
and $|\partial_a|$ is the length of clause $a$ (initially $|\partial_a|=k$).
\end{comment}
For random $k$-SAT problems, it appears discontinuously at the clustering threshold $\alpha_d$ for $k \ge 4$ \cite{montanari-clusters-2008} in the large $N$ limit, and then decreasing with increasing $\alpha$ until the condensation transition at $\alpha_c$ where it becomes null. Thus, for random $k$-SAT formulas in the rigidity phase, we expect a larger complexity to possibly correspond to an easier problem. This is confirmed by observation of fairly monotonic decrease in the number of violations as a function of initial complexity as  shown in Fig.~\ref{fig:complex}.

\begin{figure}
    \centering
    \includegraphics[width=\columnwidth]{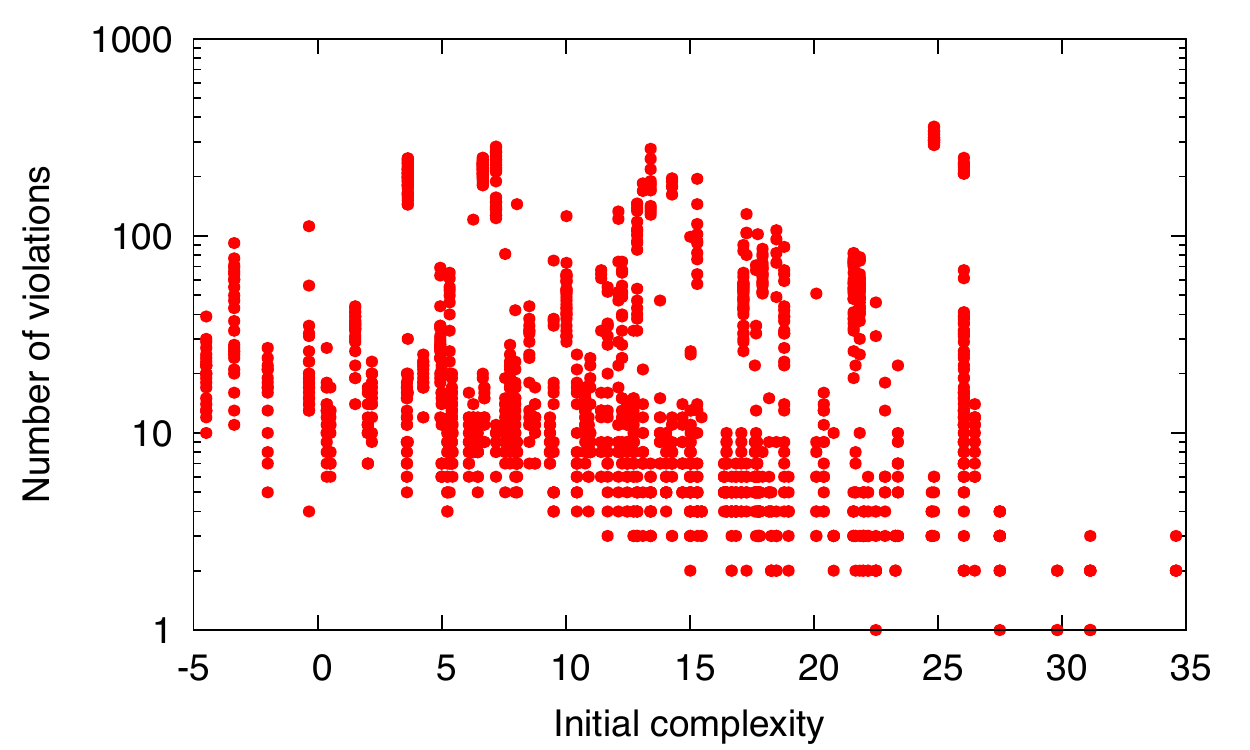}
    \caption{The energy or number of violations reached by every BSP run on the 100 instances of random 4-SAT studied. Instances are sorted according to the initial complexity which correlates well with the instances hardness (indeed the lowest energy decreases on average with increasing complexity). We notice that for many instances and without any evident correlation with the instance hardness, the BSP can often get trapped in configurations of very high energy, up to 2 order of magnitude larger than the optimum.}
    \label{fig:manyRuns}
\end{figure}

Hereafter, we use the initial complexity as a proxy for the instance hardness. We plot in Fig.~\ref{fig:manyRuns} the energy or number of violations reached by each BSP repetition on the 100 instances of random 4-SAT studied. Instances are sorted according to the initial complexity which correlates well with the instances hardness as explained above. Indeed, except a few exceptions, the lowest energy decreases  with increasing complexity. We notice that for many instances, and without any evident correlation with the instance hardness, the BSP can often get trapped in configurations of very high energy, up to 2 order of magnitude larger than the optimum. This is a very delicate point which was not noticed before, because the BSP algorithm were used for instances at large $N$ limit with sufficiently less constraints which corresponded to a smaller $\alpha$ value well before the estimated rigidity threshold.  Fig.~\ref{fig:manyRuns} is alarming since it indicates that the behavior of BSP is far from being deterministic (on this sizes), at variance with its precedent algorithm SP, and it has high volatility and unreliable performance across various runs for various instances. Thus, we need to estimate the number of repetitions required to ensure convergence to the optimal energy with a reasonable certainty. Moreover in many instances we observe a fairly large gap between the high energy local minima reached by the majority of runs and the low energy minima: this gap is particularly significant as it preclude a smooth convergence towards the optimal minima by increasing the number of repetitions. In other words, the behavior of the BSP algorithm seems to undergo a kind of first order transition between a poorly informative fixed point to a highly informative one. This behavior has been observed in many high-dimensional inference problems \cite{zdeborova2016statistical} and it is a the basis of the algorithmic gap in many hard inference problems \cite{gamarnik2021overlap}.

\begin{figure}
    \centering
    \includegraphics[width=\columnwidth]{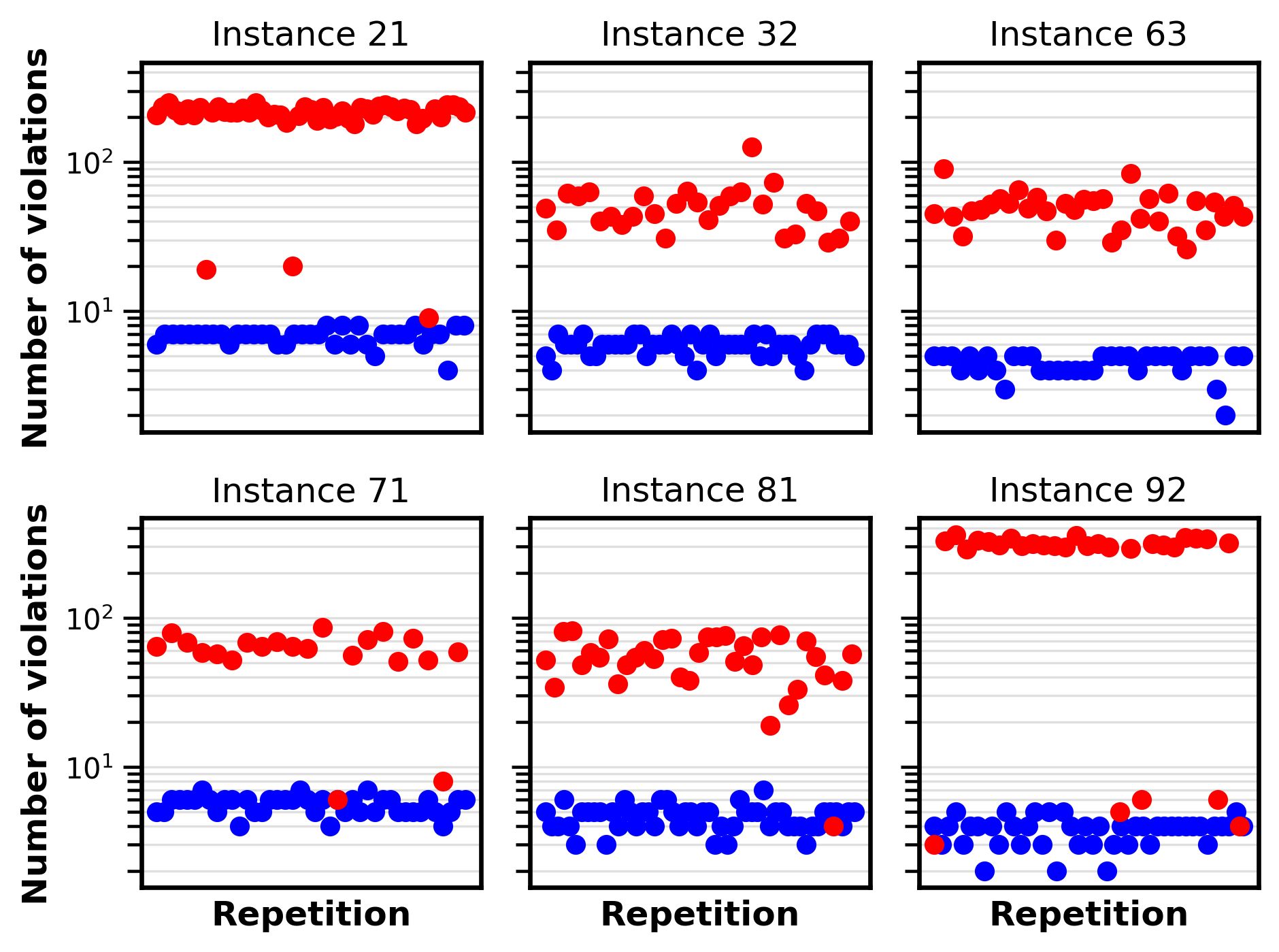}
    \caption{Six sample instances of random 4-SAT where fluctuations in the output of BSP are severe with significant gaps to the solutions found by NMC for most or all repetitions. Each red/blue point corresponds to number of violations obtained in a single repetition of the BSP/NMC algorithm. In two instances BSP fails in finding any close-to-optimal configuration. The instances are indexed according to their initial complexity, so large fluctuations for BSP persist across instances with very different number of clusters of solutions.  This implies small success probability of obtaining sufficiently  low-energy states for BSP as shown in Fig. \ref{fig:sucessrate}.}
    \label{fig:BSPfluct}
\end{figure}

Having understood the existence of large energy fluctuations in the output of the BSP algorithm, we would like to quantify how often the BSP algorithm can report a very bad configuration, i.e.\ a configuration of very high energy. In Fig.~\ref{fig:BSPfluct} we show data from 4 different instances where fluctuations look severe (left and central panels), plus 2 instances where BSP failed to find close-to-optimal configurations (right panels). Each red data point corresponds to the energy or number of violations obtained in a single BSP run. For each instance there are many runs obtained with a total running time of 5 hours. We observe that the majority or even the totality of runs returns a very poor configuration, whose energy is higher by at least one order of magnitude, and even more, with respect to the optimal one obtained via NMC (lower blue dots). Only rarely BSP is able to ``break the ceiling'' and enter the region of very low energy configurations. 

\begin{figure}
    \centering
    \includegraphics[width=\columnwidth]{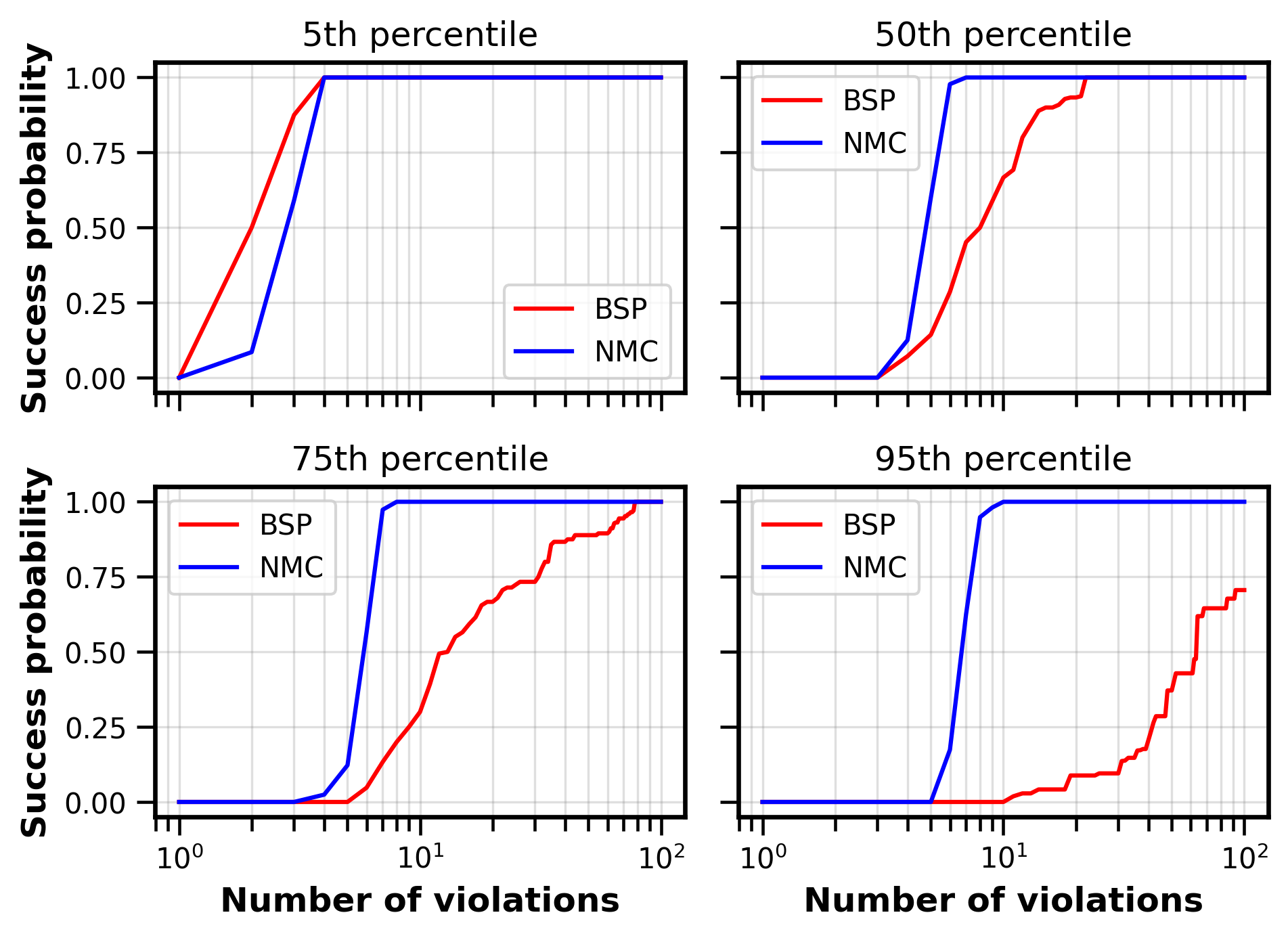}
    \caption{The success probability of obtaining a given quality of solution, spanning two orders of magnitude, for BSP and NMC solvers on four distinct instances. We see that BSP has a higher success rate on top 5\% easiest instances. For median instances the performance of BSP and NMC are fairly close. However for 75\% or higher percentile instances, and in particular top 5\% hardest instances, obtaining very low number of violations with BSP becomes very hard to achieve, while NMC still can guarantee single digit violations in almost every run.}
    \label{fig:sucessrate}
\end{figure}

What is even more impressive in Fig.~\ref{fig:BSPfluct} is that all runs of NMC return consistently high-quality solutions which even at worse cases outperform BSP runs. This means the output of NMC is extremely robust and stable. We expected this behavior in Monte Carlo based algorithms, but here the energy landscape is full of local minima at high energies as evident from BSP highly fluctuating performance and WalkSAT consistently poor performance, shown by grey dots in Fig.~\ref{fig:BSPfluct}). Consequently, consistently reaching the lowest minima is a non-trivial result for NMC implying the ability to bypass or penetrate though tall energy barriers that are hindering other solvers. In Fig.~\ref{fig:sucessrate}, we compare the success rate of NMC and BSP for a wide spectrum of instances from easiest 5th percentile to hardest 95th percentile showing a commanding advantage for NMC. For example, in order to reach $O(10^{-4})$ approximation ratio (i.e.\ around 5-10 violated clauses) for top 25\% hardest instances, NMC is doing very well with a success probability close to 1, while BSP has essentially a success probability very close to zero.

In summary, running BSP for a finite (and small) number of repetitions looks to have a serious flaw for hardest random 4-SAT instances in the rigidity phase. With high probability the algorithm can get stuck in the high energy minima and misses the optimal configurations. This is due to the fact that the BSP algorithm is not able to decimate enough variables for such instances and WalkSAT cannot handle too large residual formula leading to very poor solutions. How this phenomenon will change by increasing the size of the formulas is out of the scope of the present paper and will be presented in a forthcoming study. Nonetheless the appearance of this kind of problem on formulas comparable in size to realistic application is of considerable importance. We expect all these problems to arise when running BSP on non-random SAT instances, as industrial instances are.
In all such applications, we anticipate that the use of NMC is preferred by a large margin to SP and BSP.

%%%%%%%%%%%%%%%%%%%%%%%%%%%%%%%%%%%%%%%%%%%%%%%%%%%%%%%%%%%%%%%%%%%%%%%%%
\begin{figure}[t]
    \centering
    \includegraphics[width=\columnwidth]{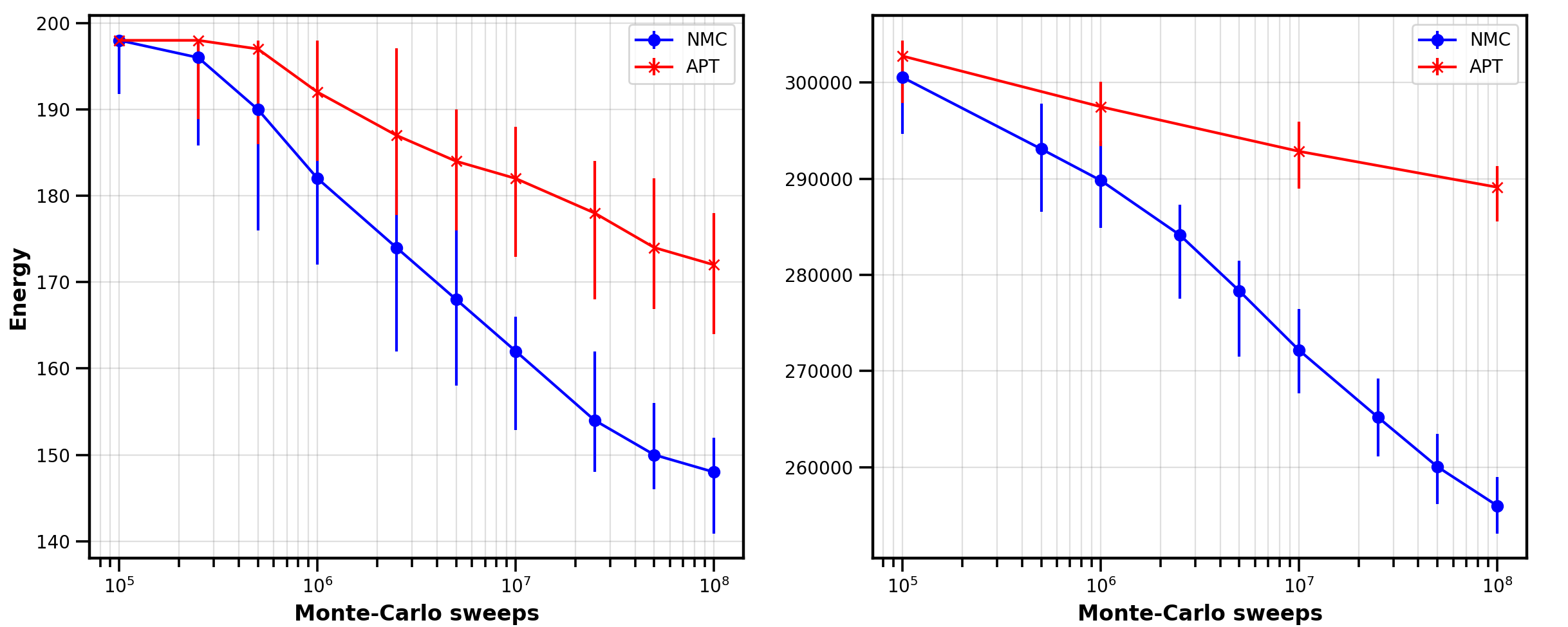}
    \caption{Running time of APT and NMC for two industry instances from QAPLIB \cite{QAPLIB_website, burkard_QAP_1998}, each containing a significant amount of structure residing on a fully connected graph: Esc32a (left), with 1032 binary variables and over one million interactions, and Tho40 (right) with 1600 binary variables and about 2.5 million interactions. The median of residual energy and 95\% and 5\% quantiles are plotted from 50 repetitions as a function of MC sweeps. We observe that NMC obtains much higher quality of solutions not only on the median but also on its worst runs and thus  provides a robust output, even though it is inherently a nonequilibrium process. Each point in these plots is optimized with Vizier as illustrated in Fig. \ref{fig:Vizier_sc32a_tho40}.}
    \label{fig:sc32a_tho40}
\end{figure}
%%%%%%%%%%%%%%%%%%%%%%%%%%%%%%%%%%%%%%%%%%%%%%%%5%%%%%%%%%%%%%%%%%%%%%%%%%

\section{Industrial QAP problems}
\label{App:Industrial_QAP}

%%%%%%%%%%%%%%%%%%%%%%%%%%%%%%%%%%%%%%%%%%%%%%%%%%%%%%%%%%%%%%%%%%%%%%%%%
\begin{figure}[ht]
    \centering
    \includegraphics[width=\columnwidth]{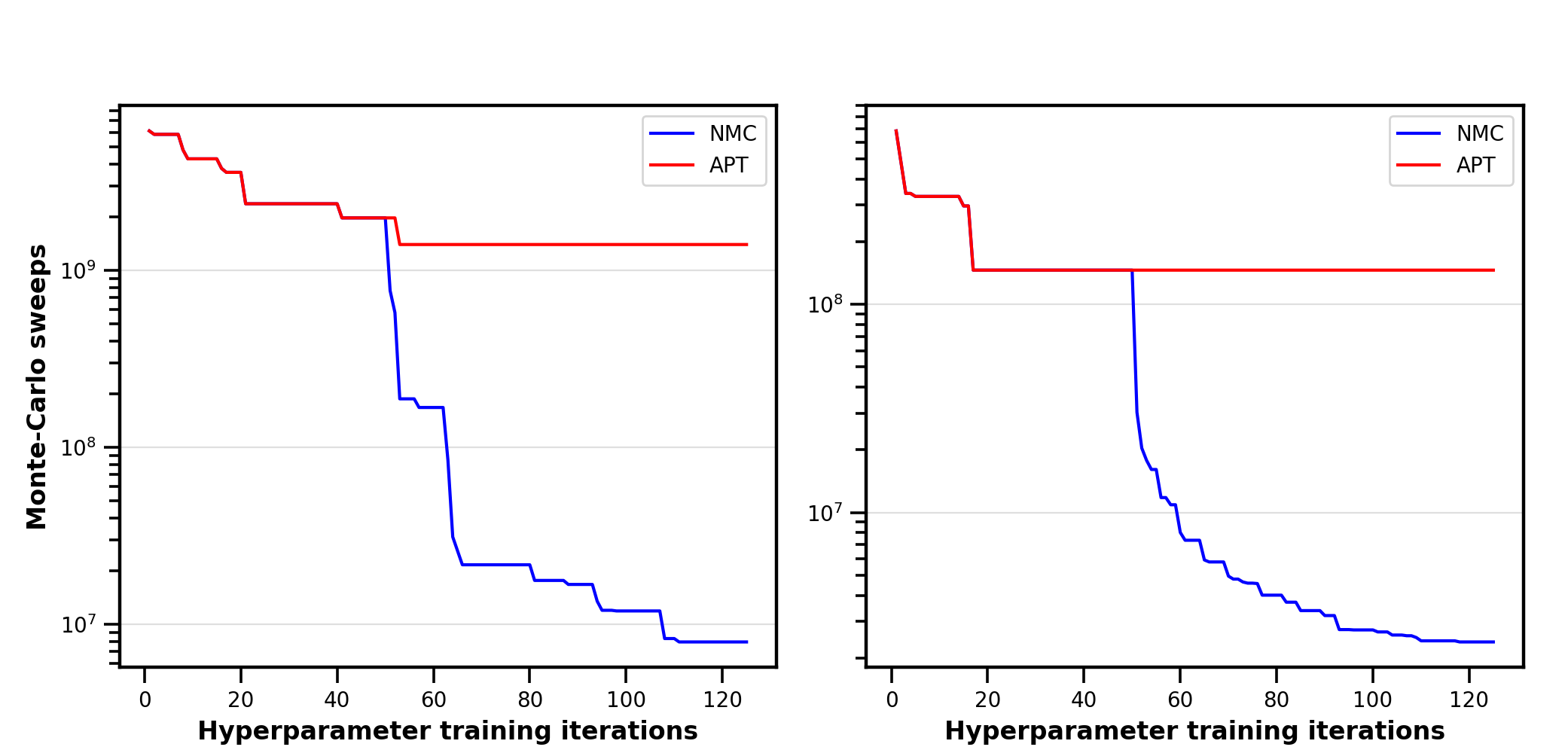}
    \caption{The Vizier hyper-parameter optimization trials for APT and NMC for two industrial QAP instances Esc32a (left) and Tho40 (right) with sizes 1032 and 1600 binary variables respectively (from QAPLIB \cite{QAPLIB_website, burkard_QAP_1998}). We observe that the nonlocal strategy provides about 2 orders of magnitude speedup to arrive at a target approximate solution of $10^{-2}$ and can find such optimal hyper-parameters with about 100 trials. In contrast, APT performance saturates in about 50 trials and cannot be improved further even with thousands of trials, thus never reaching the fast timescales that are available to the nonlocal strategy.}
    \label{fig:Vizier_sc32a_tho40}
\end{figure}
%%%%%%%%%%%%%%%%%%%%%%%%%%%%%%%%%%%%%%%%%%%%%%%%5%%%%%%%%%%%%%%%%%%%%%%%%%

We investigated the performance of our nonlocal NMC against local APT on some of the industrial instances from QAPLIB \cite{burkard_qaplib_1997}. 
These instances have either known optimal solutions or empirical best feasible solutions with tight lower bounds. For a majority of the instances in the Esc32 class, using our benchmarking framework for APT on Google's distributed computing platform, we could complete 50 repetitions of millions of MC sweeps, in less than 1 min and 30 seconds to find their optimal solutions (with a total number of sweeps for each run ranging from $10^4$ to $10^6$). In Fig. \ref{fig:sc32a_tho40}, we show the cost for the median, 5\% and 95\% percentile runs from 50 repetitions on two of the hardest instances: Esc32a which contains 1032 binary variables and over one million interactions, and instance Tho40 involving 1600 binary variables and 2.5 million interacting terms. The optimal solution for Tho40 is not known and the best feasible solution is actually obtained with simulated annealing \cite{QAPLIB_website}, which is generally believed to be suboptimal compared to the APT algorithm for sufficiently hard problem classes. We observe that, by increasing the MC sweeps, the NMC penetrates significantly lower energy levels such that its worse runs completely outperform the APT best runs out of 50 repetitions. The wall-clock time to complete total $10^8$ sweeps for APT over a couple of cores for Esc32a and Tho40 were about 30 minutes and 3 hours respectively. The overhead of nonlocal moves including LBP runs amounts to 5\% to 20\% on various QAP instances but can be significantly reduced. For each point in these plots, we have used the Vizier hyper-parameter optimization as demonstrated in Fig. \ref{fig:Vizier_sc32a_tho40}.

We note that the Vizier could find in less than 100 trials a set of hyperparameters for NMC that leads to more than 2 orders of magnitude speedup for approximating these instances. However, Vizier could not improve the running time of APT even with thousands of trials to achieve anywhere near the NMC performance. We expect that the speedup observed here over APT will increase with the size and hardness of industrial QAP instances.

\end{document}